\documentclass[journal]{IEEEtran}
\ifCLASSINFOpdf
\else
\fi

\usepackage{mathrsfs}
\usepackage{amsmath}
\usepackage{amsfonts}
\usepackage{amsthm}
\usepackage{amssymb}
\usepackage{graphicx}
\usepackage{subfigure}
\usepackage{indentfirst}
\usepackage{array}
\usepackage{epstopdf}
\usepackage{cite}

\usepackage{threeparttable}
\usepackage{adjustbox}
\usepackage{tabularx}
\usepackage{booktabs}

\usepackage{enumerate}
\usepackage{bm}
\usepackage{dsfont}
\usepackage{multirow}
\usepackage{verbatim}
\usepackage{stfloats}
\usepackage{makecell}
\usepackage{cancel}
\usepackage[dvipsnames]{xcolor}
\usepackage[colorlinks=true, linkcolor=blue, citecolor=blue]{hyperref}

\usepackage{algorithm}
\usepackage{algorithmicx}
\usepackage[noend]{algpseudocode}

\algnewcommand{\LineComment}[1]{\State {\color{blue}\(\triangleright\) #1}}

\usepackage{xcolor}
\usepackage{color}
\usepackage{amssymb}

\newcommand{\haty}{\hat{\bm{y}}}
\newcommand{\hatym}{\haty_{\mathrm{M}}}
\newcommand{\checky}{\check{\bm{y}}}
\newcommand{\checkx}{\check{\bm{x}}}
\newcommand{\y}{\bm{y}}
\newcommand{\hatx}{\hat{\bm{x}}}
\newcommand{\x}{\bm{x}}

\begin{document}

    \title{ResiComp: Loss-Resilient Image Compression via Dual-Functional Masked Visual Token Modeling}

    \author{Sixian Wang,~\IEEEmembership{Member,~IEEE},
	Jincheng Dai,~\IEEEmembership{Member, IEEE},
	Xiaoqi Qin,~\IEEEmembership{Senior Member, IEEE},\\
	Ke Yang,~\IEEEmembership{Graduate Student Member, IEEE},
	Kai Niu,~\IEEEmembership{Member, IEEE},
	and Ping Zhang,~\IEEEmembership{Fellow, IEEE}
	
	\thanks{This work was supported in part by the National Key Research and Development Program of China under Grant 2024YFF0509700, in part by the National Natural Science Foundation of China under Grant 62371063, Grant 62293481, Grant 62321001, and Grant 92267301, in part by the Beijing Municipal Natural Science Foundation under Grant L232047 and Grant 4222012, in part by Young Elite Scientists Sponsorship Ph.D. Program by CAST, in part by Program for Youth Innovative Research Team of BUPT under Grant 2023YQTD02. \emph{(Corresponding author: Jincheng Dai)}}
	
	\thanks{Sixian Wang, Jincheng Dai, Ke Yang, and Kai Niu are with the Key Laboratory of Universal Wireless Communications, Ministry of Education, Beijing University of Posts and Telecommunications, Beijing 100876, China. (Email: daijincheng@bupt.edu.cn, sixian@bupt.edu.cn)}
	
	\thanks{Xiaoqi Qin and Ping Zhang are with the State Key Laboratory of Networking and Switching Technology, Beijing University of Posts and Telecommunications, Beijing 100876, China.}
	
	\thanks{Code and pretrained models are available at \url{https://github.com/wsxtyrdd/ResiComp}.}
	
	\vspace{-1em}
	}

    \maketitle

    \begin{abstract}
        Recent advancements in neural image codecs (NICs) are of significant compression performance, but limited attention has been paid to their error resilience.
        These resulting NICs tend to be sensitive to packet losses, which are prevalent in real-time communications.
        In this paper, we investigate how to elevate the resilience ability of NICs to combat packet losses.
        We propose \emph{ResiComp}, a pioneering neural image compression framework with feature-domain packet loss concealment (PLC).
        Motivated by the inherent consistency between generation and compression, we advocate merging the tasks of entropy modeling and PLC into a unified framework focused on latent space context modeling.
        To this end, we take inspiration from the impressive generative capabilities of large language models (LLMs), particularly the recent advances of masked visual token modeling (MVTM).
        In specific, \emph{ResiComp} develops a bi-directional masked Transformer to model the contextual dependencies among latents with dual-functionality: 1) iteratively acts as a conditional entropy model to boost compression efficiency; 2) operates latent PLC to improve resilience.
        During training, we integrate MVTM to mirror the effects of packet loss, enabling a dual-functional Transformer to restore the masked latents by predicting their missing values and conditional probability mass functions.
        Our \emph{ResiComp} jointly optimizes compression efficiency and loss resilience.
        Moreover, \emph{ResiComp} provides flexible coding modes, allowing for explicitly adjusting the efficiency-resilience trade-off in response to varying Internet or wireless network conditions.
        Extensive experiments demonstrate that \emph{ResiComp} can significantly enhance the NIC's resilience against packet losses, while exhibits a worthy trade-off between compression efficiency and packet loss resilience.
        Additionally, packet-level simulations, conducted using diverse network models based on real traces, demonstrate that \emph{ResiComp} exhibits much better robustness to fluctuating network conditions compared to redundancy-based approaches like VTM + FEC.
    \end{abstract}

    \IEEEpeerreviewmaketitle

    \section{Introduction}\label{section_introduction}

	\subsection{Background}
    \IEEEPARstart{R}{eal-time} visual applications have been recognized as one of the biggest challenges according to the latest video developer report~\cite{bitmovin}, primarily attributed to the incompatibility between highly variable network conditions and increasing user demands for a better quality of experience (QoE).
    When packet loss or errors occur during data transmission, the general approach is to request retransmission of the affected packets.
    However, retransmission is suitable only for scenarios with short round trip times (RTTs) due to the strict latency requirement of real-time communications (RTC) applications.
    For all other cases, lost packets may not be concealed via retransmission, especially in real-time video systems which hope frames to be played as soon as they are decoded, such as FaceTime, Skype, and WebRTC~\cite{WebRTC}.
    Resending any packets can significantly contribute to the overall delay experienced by the user, leading to poor user QoE\@.

    For this reason, another broad viable solution employs forward error correction (FEC) codes to recover lost packets within an acceptable latency.
    With interpacket FEC codes, one can add at least $r$ redundancy packets (also termed ``parity'' packets) in advance to cover an expected no more than $r$ lost packets before transmission.
    Since the instant packet loss ratio cannot be known in advance, too much redundancy degrades the compression efficiency, and inadequate protection still needs for retransmission until enough packets are received.
    In response, most RTC systems dynamically adjust FEC redundancy based on current network conditions -- increasing redundancy after packet loss has already affected some frames and scaling it back after the packet loss has subsided.
    However, this approach is highly inefficient, as the adjustment always lags behind the ever-fluctuating network conditions.
    When we look back at the evolution of classical image/video codecs (from H.261~\cite{bhaskaran1997image} in 1988 to H.266~\cite{VVC} in 2020), it is evident that remarkable progress has been achieved for better compression efficiency.
    Nevertheless, there remains considerable room for enhancement in resilience aspects.
    Contemporary video systems still rely on a hybrid retransmission and FEC codes to handle potential packet losses, which heavily depends on the accurate and timely estimation of network conditions, particularly the number of parity packets $r$.

    In the case of emerging neural image codecs (NICs)~\cite{balle2020nonlinear}, this issue becomes more critical.
    NIC has achieved significant advancements, surpassing traditional image codecs like BPG and VVC-intra in compression efficiency, but its resilience to packet loss remains largely unexplored yet.
    NICs function by encoding an image into a sequence of latents with nonlinear transform networks and then models the latents' marginal probability in an automatically learned entropy model.
    Towards better rate-distortion performance, various nonlinear transform networks~\cite{cheng2020learned,  zou2022devil, he2022elic, lu2022high, liu2023learned} and entropy models have been proposed~\cite{minnen2018, minnen2020channel, he2021checkerboard, jiang2023mlic}.
    However, these neural compression networks primarily trained for source compression are very sensitive to packet losses, which are common in RTC applications but overlooked in the past.
    With the increasing demand of RTC services, it is the very time to investigate the loss resilient capability of NICs.

	\subsection{Challenges in Building Loss-Resilient NICs}
    In this paper, we aim to build a neural compression framework that combines high efficiency and inherent resilience to various packet loss rates.
    Developing such a neural compression framework poses two fundamental questions.

    Firstly, how to effectively conceal latent damage caused by packet loss?
    NICs, essentially variational autoencoders, have shown susceptibility to various perturbations and backdoor attacks~\cite{chen2023towards, yu2023backdoor}.
    Packet loss, particularly at the bottleneck layer, results in permanent damage to latent representations.
    If not managed properly, these impaired latents can propagate to the entire image, reducing the overall image reconstruction quality, which differs from classical image codecs using block-wise linear transform.
    As a result, it is more effective for NICs to conceal the latents loss in the feature-domain, rather than employing the RGB-domain post-processing tools widely-used in classical resilient video coding~\cite{wang2000error}.


    Secondly, how to achieve a better trade-off between resilience and efficiency while limiting error propagation?
    NICs utilize entropy codecs (e.g., arithmetic coding) to losslessly encode groups of latents into the bitstrings of packets, which are vulnerable to transmission errors.
    Even a single bit error can cause the NIC decoder losing synchronization~\cite{balle2018integer, tian2023towards}, rendering the successive received bitstrings in the same packet useless.
    Moreover, advanced NICs employ context entropy models~\cite{minnen2018, he2021checkerboard, minnen2020channel, he2022elic, lu2022high} for more accurate density estimation, which further exacerbates the error propagation.
    Errors in packets containing conditioning latents can affect all dependent latents, even those encoded in other received packets.
    To limit the extend of error propagation, resilient codecs primarily focus on detecting, correcting, or concealing errors, but this requires less coding dependencies among latents which may degrade their compression efficiency~\cite{wang1998error}.
    Hence, there exists a natural trade-off between achieving better compression efficiency and ensuring error resilience, this is also conceptually aligned with the fundamental trade-off between efficiency and reliability in Shannon information theory.
	
	\subsection{ResiComp: A Loss-Resilient Compression Framework}
	
	\begin{figure}[ht]
		\setlength{\abovecaptionskip}{0.cm}
		\setlength{\belowcaptionskip}{-0.cm}
		\centering
		\includegraphics[scale=0.35]{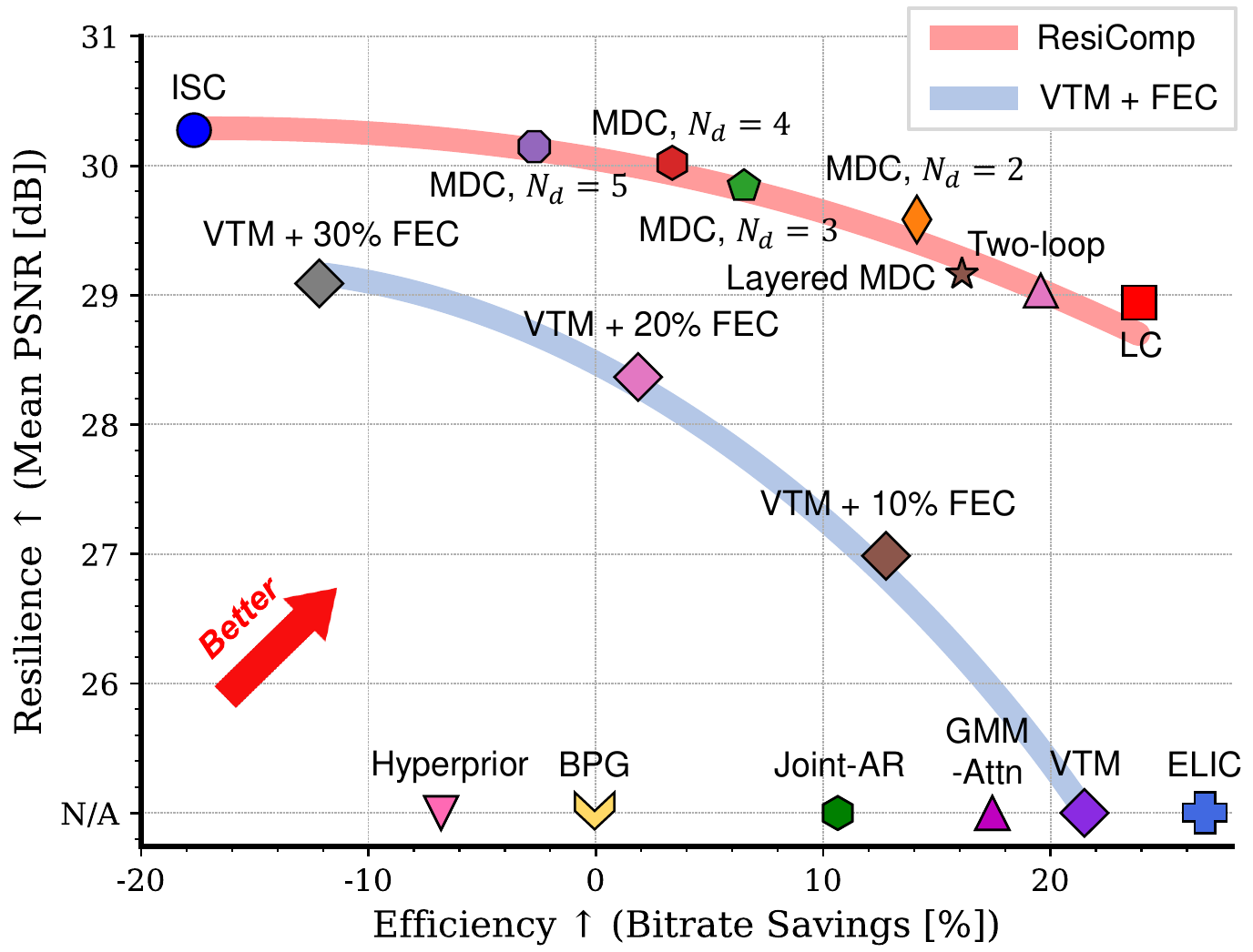}
		\caption{Efficiency-resilience trade-off on Kodak.  Higher positions in the top-right corner indicate better performance. The resilience performance of some codecs is marked as "N/A," indicating that due to frequent decoding failures, their average PSNR falls below 25 dB.
			The detailed description will be presented in Section \ref{resilience-comparison}.}
		\label{Fig_benchmark}
	\end{figure}
	
    To tackle the aforementioned two questions, we propose to integrate the context modeling capability of large language models (LLMs) into loss-resilient compression.
    This idea is inspired from the intrinsic consistence between generation and compression, a principle long established in these two fields.
    Deep generative models can also be transformed into lossless compressors and vice versa~\cite{jakub2022deep}.
    Benefited from the impressive generative capability of LLMs, very recent advances in NICs~\cite{mentzer2023m2t, el2022pqmim} have exploited masked visual token modeling (MVTM) in lossy compression and achieve remarkable rate-distortion results.
    Motivated by this, we aim to bridge entropy modeling and feature-domain {\color{blue} packet loss concealment (PLC)} with the strong context modeling capability from MVTM\@.

    Specifically, we propose \emph{ResiComp}, the first neural image compression framework that jointly optimizes for compression efficiency and resilience.
    \emph{ResiComp} segments latents into multiple slices with a presupposed pattern, and each slice will be entropy encoded into bitstream using (conditional) probability mass function (PMF) and packetlized independently.
    In this way, the extend of error propagation across packets is limited, equating the impact of packet loss to the random masking of latent slices.
    Building on this equivalence, \emph{ResiComp} develops a bi-directional Transformer backbone with dual-functionality: as a conditional entropy model to enhance compression efficiency or as a feature-domain PLC to improve resilience.
    We integrate MVTM during the training phase, empowering the Transformer learn to reconstruct masked latents by predicting both their missing values (via the PLC head) and conditional PMFs (via the density head).
    The total loss consists of three parts: the estimated rate loss for masked tokens and two distortion losses for reconstructions from quantized and recovered tokens.
    At the inference stage, the bi-directional Transformer serves a dual-functionality.
    It can iteratively function as a conditional entropy model to enhance compression efficiency or act as a feature-domain PLC to improve resilience.
    With the dual-functional context modeling, the error resilience part is optimized jointly with efficiency part to push the boundary~further.

    Based on the flexible dual-functional context modeling paradigm, we further propose the idea of \emph{context mode} to explicitly control the efficiency-resilience trade-offs.
    A context mode manages whether each latent slice should be encoded in \emph{intramode} (without referencing previous latents) or \emph{intermode} (with context modeling).
    By scheduling the usage of contextual dependencies, \emph{ResiComp} can achieve multiple efficiency-resilience trade-off adapted to network conditions within a single model.
    Comparing with redundancy-based schemes, we show \emph{ResiComp} is more robust to the changing network characteristics, and reduces the decoding failure ratio at lower bandwidth cost.
    A preview comparison between our \emph{ResiComp} and VTM + FEC is shown in Fig. \ref{Fig_benchmark}.

    The main contributions of this paper are listed as follows.
	\begin{enumerate}
		\item \emph{Loss-resilient compression framework}: Inspired by inherent consistence between generation and compression, we propose \emph{ResiComp}, the first loss-resilient neural image compression framework that jointly optimizes all the components for efficiency-resilience trade-off.
		\item \emph{Dual-functional context modeling}: We propose to merge the entropy modeling and feature-domain PLC into a unified approach focused on latent space context modeling.
		By incorporating MVTM training, we develop a dual-functional Transformer to achieve flexible entropy modeling and error concealment with one model.
		\item \emph{Adaptive coding mode scheduling}: Based on the flexible context modeling paradigm, we further propose the idea of \emph{context mode} to explicitly scheduling the spatial dependencies within a single model.
		In this way, \emph{ResiComp} can adjust itself to multiple efficiency-resilience trade-offs based on the network quality.
	\end{enumerate}

    The remainder of this paper is organized as follows.
    In Section~\ref{section_related_work}, we review related works on neural image compression and loss resilience techniques.
	The proposed dual-functional masked token modeling method is introduced in Section~\ref{section_proposed_method}, while Section~\ref{sec:loss-resilient-image-compression} details the process of compression, packetization, and transmission. The experimental results on various datasets and network models are discussed in Section~\ref{section_experiments}. Finally, Section~\ref{section_conclusion} concludes the paper.

    \section{Background and Related Work}\label{section_related_work}


    \subsection{Lossy Image Compression}\label{subsec:lossy-image-compression}

    Lossy image codecs are primarily based on the paradigm of \emph{transform coding}~\cite{trans_coding}, whose encoding process consists of three steps: decorrelation, quantization, and entropy coding.
    In principle, they aim to search for the \emph{optimal compact representation} of the input source in a computationally feasible way that leads to the best rate-distortion (RD) performance~\cite{rd_theory} defined as
    \begin{equation}
    \label{RDO}
    J = R + \lambda D,
    \end{equation}
    where $\lambda$ denotes the Lagrange multiplier that controls the desired compression trade-off between the rate and distortion. The bit rate term $R$ represents the average number of bits needed to encode the input data, and the distortion term $D$ assesses the similarity between the input and its reconstruction.

    \subsubsection{Classical Image Codecs}

    In conventional image codecs such as JPEG, JPEG2000, BPG\cite{BPG}, and VVC intra\cite{VVC_intra}, the transforms are typically \emph{linear and invertible} such as discrete cosine transform (DCT) and discrete wavelet transform (DWT). To improve RD performance, these \emph{linear transform coding} schemes often employ continuously expanded available coding modes to search the best predictive manner for reducing spatial redundancies, whose transform, quantizer, and entropy code are separately optimized through manual parameter adjustment.

    \subsubsection{Neural Image Codecs}
    Most neural image codecs (NICs) are based on \emph{nonlinear transform coding}~\cite{balle2020nonlinear}, which employs deep neural networks (DNNs) to implement various components and learns them end-to-end on the data of interest.
    
	In general, a NIC encodes an image $\x$ into compact (ideally decorrelated) transform coefficients $\y$ using an \emph{analysis transform} $g_a$. These coefficients are then scalar quantized element-wise to obtain a discretized representation $\haty = Q(\y)$. With an entropy model $p_{\haty}$, the discrete symbols $\haty$ are converted into a bitstring with the expected length $-\mathbb{E}[\log p_{\haty}(\haty)]$ via an entropy encoder. 
    Under the reliable transmission promise, the receiver recovers $\haty$ losslessly and reconstructs the original image using a \emph{synthesis transform} $g_s$.	
    
    State-of-the-art NICs~\cite{cheng2020learned, he2022elic, zou2022devil, lu2022high} are already surpassing the advanced traditional method VVC, mainly due to more efficient nonlinear transforms and expressive entropy models. 
    Specifically, the efficient nonlinear transform blocks explored in existing works include residual blocks~\cite{cheng2020learned, he2022elic}, vision transformers~\cite{zou2022devil}, and their combinations~\cite{lu2022high, liu2023learned, jiang2023mlic}.
    As for the entropy models, Ballé \emph{et al.} \cite{balle2016} first proposed a non-adaptive, fully-factorized entropy model to approximate $p_{\haty}$, later extending it to the hierarchical form~\cite{balle2018}.
    To achieve more accurate and efficient density estimation, advanced NICs encode quantized latents with $L$ groups, and factorize the density model to be a joint hierarchical and group-based auto-regressive form: $\prod_{i=1}^{L} p(\haty_{i} | \haty_{<i}, \hat{\bm{z}})$ \cite{minnen2018, lu2022high, he2022elic}, where $\hat{\bm{z}}$ denotes the hyperprior \cite{balle2018}. 
    Various contextual dependencies have been explored, including spatial context models~\cite{minnen2018, he2021checkerboard, mentzer2023m2t}, channel-wise models~\cite{minnen2020channel}, and hybrid spatial-channel models~\cite{he2022elic, lu2022high}.
	
    However, existing NICs, primarily designed for source compression, produce fragile bitstreams that are susceptible to perturbations, such as bit errors~\cite{balle2018integer, tian2023towards}, which hinder its applications to RTC scenarios. 
    In this work, we investigate how to elevate the resilience of NICs for packet losses.

    \subsection{Loss-Resilient Techniques}\label{subsec:loss-resilience-techniques}
    To transmit data packets over unreliable networks, various techniques have been developed~\cite{wang2000error} to detect, correct or conceal errors, but they also intentionally make the source coder less efficient than it can be.
    The most representative schemes including forward error correction (FEC) and packet error concealment (PLC).

    \subsubsection{Forward error correction (FEC)}
    FEC is a basic method used to protect compressed bitstreams from transmission errors.
    It works by adding redundancy to the data at the sender's side, either at the application or transport layer.
    Specifically, FEC encodes $N_{k}$ data packets and adds $N_{r}$ parity packets, so that the original data packets can be recovered if any subset of $N_{k} (1 + \rho)$ packets out of the total $(N_{k} + N_{r})$ packets are received \cite{4427233}.
    Here $\rho$ denotes the proportion of additional redundancy packets required for reconstruction, and an ideal FEC code requires no decoder overhead, i.e., $\rho = 0$.
    Common error-resilient channel coding methods include Reed-Solomon codes, low-density parity check (LDPC) codes, fountain or rateless codes, etc.
    However, the significant challenge of FEC is to decide the right number of parity packets.
    It is tricky since the exact number of loss packets $N_{l}$, can never be known in advance.
    When $N_{l} > N_{r}$, the redundancy will be insufficient to recover lost packets, decoding fails.
    When $N_{l} < N_{r}$, the bandwidth consumed by transmitting extra $N_{r} - N_{l}$ parity packets will be wasted.
    As a result, it must adjust the redundancy based on estimation of instant link quality.
    This approach is highly inefficient to achieve a satisfactory balance between resilience and efficiency, as the adjustment always lags behind the network condition changes.

    \subsubsection{Packet loss concealment (PLC)}
    PLC techniques aim to restore the missing or delayed packets at the receiver side.
    In general, it requires the encoder to limit the extent of error propagation by splitting the data into several segments.
    Then, PLC encoder removes various redundancies (e.g., spatial, temporal, and statistical) only within each segment, which prevent the error segment from effecting other segments.
    When some packets are lost, the decoder estimates missing data in lost packets based on the received packets, using the correlations between segments in the pixel domain.
    Most classical PLC studies are tightly coupled with video coding standards, which inpaint the lost area using spatially and temporally surrounding motion vectors (MVs), neighboring pixels, or other available side information.
    Recent advances come from the learning-based image/video completion methods, which generate expressive and realistic results exploiting high-level image features.
    However, these codec-agnostic PLC tools are highly dependent on the codec's output, which limits their full potential.
    Although they are valuable at the post-processing stage of classical codecs, these pixel-domain PLC tools exhibit limited compatibility with neural codecs, since damaged latents can spread across the entire image and degrade the overall reconstruction quality.
    As a result, recent works in resilient neural speech coding jointly design feature-domain PLC with the source compression~\cite{msra_plc}.
    But in the context of resilient image coding, such a paradigm shift remains to be studied.
    This paper aims to bridge this gap, exploring the potential of addressing packet losses with feature-domain PLC in a neural image coding framework.


    \section{Dual-Functional Masked Visual Token Modeling}\label{section_proposed_method}


    \begin{figure*}[t]
        \setlength{\abovecaptionskip}{0.cm}
        \setlength{\belowcaptionskip}{-0.cm}
        \centering{\includegraphics[width=0.99\textwidth]{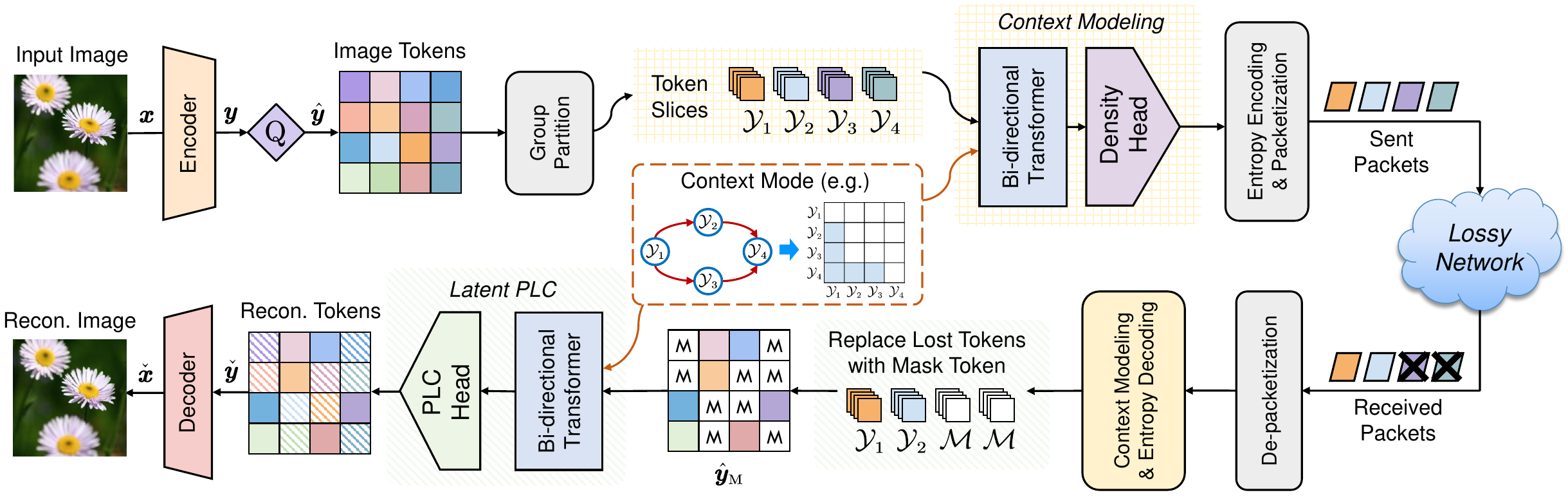}}
        \caption{
        \textbf{Overview of \emph{ResiComp}.} The encoder maps the input image into quantized image tokens. These tokens are divided into multiple slices, and each slice is entropy encoded and packetized into a packet. A bi-directional Transformer is employed with a density head for iterative entropy modeling. Some packets may be lost during transmission, and the affected image tokens are replaced with special mask tokens. We predict the masked tokens using the same Transformer, with a PLC head for packet loss concealment, and reconstruct the input image from the inpainted tokens using a decoder.
    	{The detailed loss-resilient image compression process is presented in Sec. \ref{sec:loss-resilient-image-compression}, with the workflows of the \emph{ResiComp} sender and receiver outlined in Algorithm~\ref{algorithm_sender} and Algorithm~\ref{algorithm_receiver}. The token decoding progress at the receiver is further explained in Fig. \ref{Fig_iterative_decoding}.}}
        \label{Fig_framework}
    \end{figure*}


    \subsection{Overview}\label{subsec:overview}
    \emph{ResiComp} is designed to elevate the resilience of existing neural codecs.
    To achieve this, as illustrated in Fig.~\ref{Fig_framework}, we tailor major components of an image delivery framework:
    \begin{itemize}
        \item \textbf{Tokenization:}
        The encoder $g_a(\cdot)$ transforms the input image $\bm{x}$ into a sequence of tokens $\bm{y} \in \mathbb{R} ^ {N \times C}$, comprising $N$ tokens each with a channel dimension $C$.
        \item \textbf{Quantization:}
        These tokens are then element-wise quantized to the discrete form $\haty = Q\left(g_a(\bm{x})\right)$, $Q$ denotes a standard rounding operator.
        \item \textbf{Slice Partition:}
        To transmit $\haty$ over a packet-lossy network, the partitioner divides $\haty$ into multiple token slices $\{\mathcal{Y}_{1}, \dots, \mathcal{Y}_{L}\}$ through a mapping function shared between transceiver.
        Each slice is entropy encoded and packetized into a packet.
        \item \textbf{Entropy Model:}
        It estimates the (conditional) probability mass function (PMF) of $p(\haty)$ to improve the compression efficiency.
        \emph{ResiComp} employs a flexible masked Transformer as entropy model $f_e(\cdot)$ to model $p(\haty)$ iteratively.
        \item \textbf{Context Mode Selection:}
        A context mode uniquely organizes the contextual dependencies among token slices.
        Encoding a slice with more contexts can improve the compression efficiency, but it also increases the extent of error propagation.
        \emph{ResiComp} supports diverse context modes to achieve multiple efficiency-resilience trade-offs.
        \item \textbf{Entropy Coding and Packetization:}
        The entropy coder converts each token slice $\mathcal{Y}_l$ to a bitstring in parallel, with the total expected length of $-\mathbb{E}[\log p(\haty)]$.
        Each bitstring is then packetized into a packet for transmission.
        \item \textbf{Packet Lossy Network:}
        The impact of packet losses on image tokens is shaped by multiple factors such as network conditions, packetization strategy, and the context mode utilized during encoding.
        A token slice might fail to decode for two primary reasons: (a) the loss of its bitstream; and (b) its encoding with certain contexts that, if lost, render the token slice unrecoverable.
        For simplicity, we assume that packet loss can be timely and accurately detected at the receiver end, using side information from the packet header or cyclic redundancy checking.
        Without loss of generality, we omit the process of de-packetization and entropy decoding, and model the impact of packet loss with a transfer function $\hatym = W(\haty; \phi)$, where lost tokens are replaced by a learnable masked token $\mathcal{M} \in \mathbb{R}^{1 \times C}$ ($\hatym=\haty$ in the absence of packet loss), and $\phi$ characterizes the patterns distribution of the packet loss model.
        \item \textbf{Latent PLC and Reconstruction:}
        If $\haty$ is recovered losslessly, \emph{ResiComp} generates the reconstruction $\hatx = g_s(\haty)$ with the decoder $g_s(\cdot)$, the same as other NICs do.
        In other cases,  to reconstruct from the damaged mask latents $\hatym$, \emph{ResiComp} introduces a latent PLC module $f_c(\cdot)$ to complete the lost tokens in feature-domain.
        The reconstructed tokens $\checky$ are subsequently used for generating $\checkx$ \emph{using the same decoder}, i.e., $\checkx = g_s(\checky)$.
    \end{itemize}

	\textbf{Architecture:}
	This paper focuses on how to exploit flexible contextual dependencies among tokens to achieve a better efficiency-resilience trade-off.
	For the implementation of $g_a$ and $g_s$, we adopt the ELIC autoencoder architecture with $C=192$ proposed in~\cite{he2022elic}, and the straight-through rounding operation is used to get gradients through quantization.
	For the implementation of dual-functional Transformer $f_c$, to be applicable for high-resolution inference, we employ the Swin Transformer~\cite{liu2021swin}, configured with $12$ layers, an embedding dimension of $768$ channels, a window size of $4$, a query dimension of $32$ for each head, and the an expansion layer of $4$ in each MLP layer.
	We also follow the compression-specific changes proposed in~\cite{mentzer2023m2t} for the Transformer-based entropy model: a fully connected layer is used as embedding layer for projecting ELIC bottleneck to $768$-dimensional latent space.

    \subsection{Optimization Objective}\label{subsec:optimization-objective}

	\begin{figure}[t]
		\setlength{\abovecaptionskip}{0.cm}
		\setlength{\belowcaptionskip}{-0.cm}
		\centering{\includegraphics[scale=0.39]{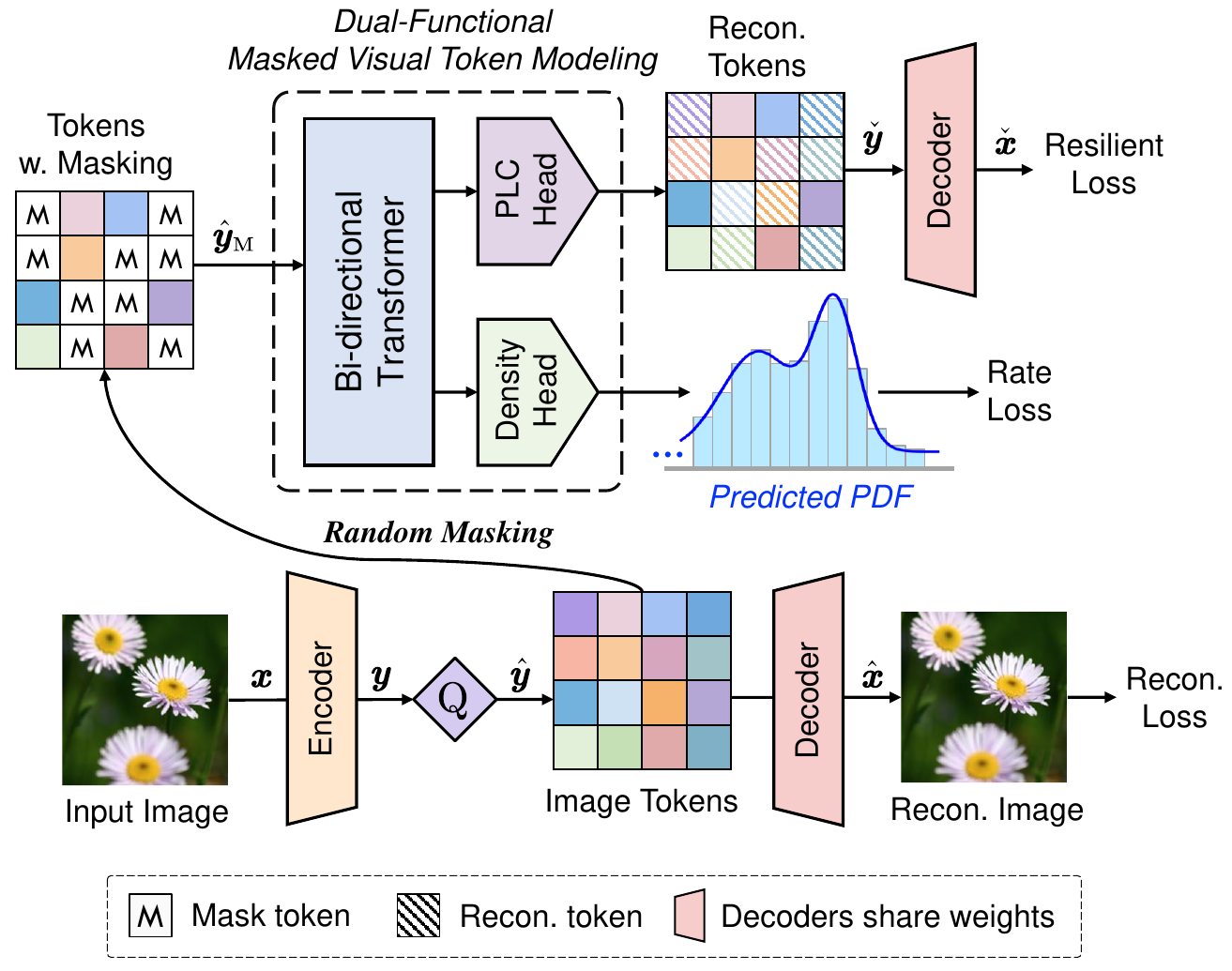}}
		\caption{\textbf{The training pipeline of \emph{ResiComp}.} We randomly sample a masking ratio from 0 to 1 and mask out tokens with $\mathcal{M}$. A bi-directional Transformer models the spatial dependencies among tokens by 1) restoring the values of masked tokens with PLC head; 2) predicting the parametric distributions of masked tokens with density head. The final loss function is the weighted sum of three terms, 1) rate loss: estimated bitrate in masking locations; 2) reconstruction loss: the reconstruct distortion from quantized tokens $\haty$; 3) resilient loss: the distortion from reconstructed tokens $\check{\bm{y}}$.}
		\label{Fig_training}
	\end{figure}

    With reliable transmission premise, one can only focus on the compression efficiency.
    Given input image distribution $p_{\bm{x}}$, a pair of nonlinear transform $g_a$ and $g_s$, and an entropy model $f_e(\cdot)$, our efficiency goal is to find the optimal trade-off between the reconstructed quality and encoding rate, expressed as:
    \begin{equation}
        \label{rate_distortion_loss}
        \mathcal{L}_{E} = \mathbb{E}_{\bm{x} \sim p_{\bm{x}}} \left[ -\log_{2}p(\haty) + \lambda d\left(\bm{x}, g_s(\haty)\right) \right],
    \end{equation}
    where the first term is the estimate bit-rate for encoding $\haty$, $d$ is the distortion function, $\lambda$ is the Lagrange multiplier that controls the desired rate-distortion trade-off.

    In the case of unreliable transmission, the resilience goal is to draw high quality reconstructions from partial packets.
    Given received masked tokens $\hatym$, and latent PLC module $f_c$, the resilience loss defined as:
    \begin{equation}
        \label{resilient_loss}
        \mathcal{L}_{R} = \mathbb{E}_{\bm{x} \sim p_{\bm{x}}} \mathbb{E}_{\hatym \sim p_{\hatym|\haty}} \left[d\left(\bm{x}, g_s(f_c(\hatym))\right) \right].
    \end{equation}

    By integrating~\eqref{rate_distortion_loss} and~\eqref{resilient_loss}, the overall loss function for \emph{ResiComp} is formulated as:
    \begin{equation}
        \label{total_loss}
        \mathcal{L} = \mathcal{L}_{E} + \alpha \cdot \mathcal{L}_{R},
    \end{equation}
    where $\alpha$ balances the scale of the two losses.
    Hereby, we seem like to achieve a joint optimization of the compression efficiency and loss resilience.

    However, we experimentally find that a pair of separately designed but jointly trained entropy model $f_e$ and latent PLC model $f_c$ does not effectively synergize (as detailed in ablation study), leading to a \emph{suboptimal} trade-off between efficiency and resilience.
    That is, the resulting autoencoder pays serious RD performance degradation in exchange for marginal resilience improvement.
    Furthermore, such improvement is closely tied to the packet loss distribution used during training, limiting their applicability in networks with high variability.

    \subsection{Dual-Functional Masked Visual Token Modeling}\label{subsec:dual-functional-masked-visual-token-modeling}

    To address such conflicts, we find a way to kill two birds with one stone.
    Particularly, we merge the two tasks within a unified approach focused on flexible context modeling.
    This approach is based on the principle that, both entropy model and latent PLC model essentially aim to leverage one group of tokens as a prior for predicting another group of tokens, with a small difference that the entropy model estimates the codewords' parameterized distributions, while the latent~PLC model directly predicts the codewords' values.
    Therefore, we propose to achieve the dual-functions with masked token~prediction, which is learned by masked visual token modeling (MVTM).
    We show the optimization objectives of efficiency and resilience are delicately aligned together in this way.

    Fig. \ref{Fig_training} illustrates the training process of proposed dual-functional MVTM\@.
    Given image tokens $\haty$, we first uniformly sample a masking ratio $r \in \left(0, 1\right)$, and randomly mask out $\lceil N \times r \rceil$ tokens in $\haty$ to place mask token $\mathcal{M}$.
    Let $\mathbf{M}$ denote a corresponding 0--1 mask which tokens are to be masked (simulating lost during transmission).
    For resulting tokens (padded with mask tokens) $\hatym$, the training objective is to reconstruct its mask locations from the unmasked tokens by predicting both values and distribution parameters.
    To achieve this, \emph{ResiComp} develops a bi-directional Transformer $f(\cdot)$, which takes $\hatym$ as input, and extracts context vectors for each mask token.
    Two MLP heads shared across tokens are concatenated after the Transformer to predict distribution parameters (via the density head $h_{\text{density}}$) and fill in the missing values (via the PLC head $h_{\text{PLC}}$), respectively.
			
	\begin{figure}[tbp]
		\setlength{\abovecaptionskip}{0.cm}
		\setlength{\belowcaptionskip}{-0.cm}
		\centering
		{\includegraphics[scale=0.41]{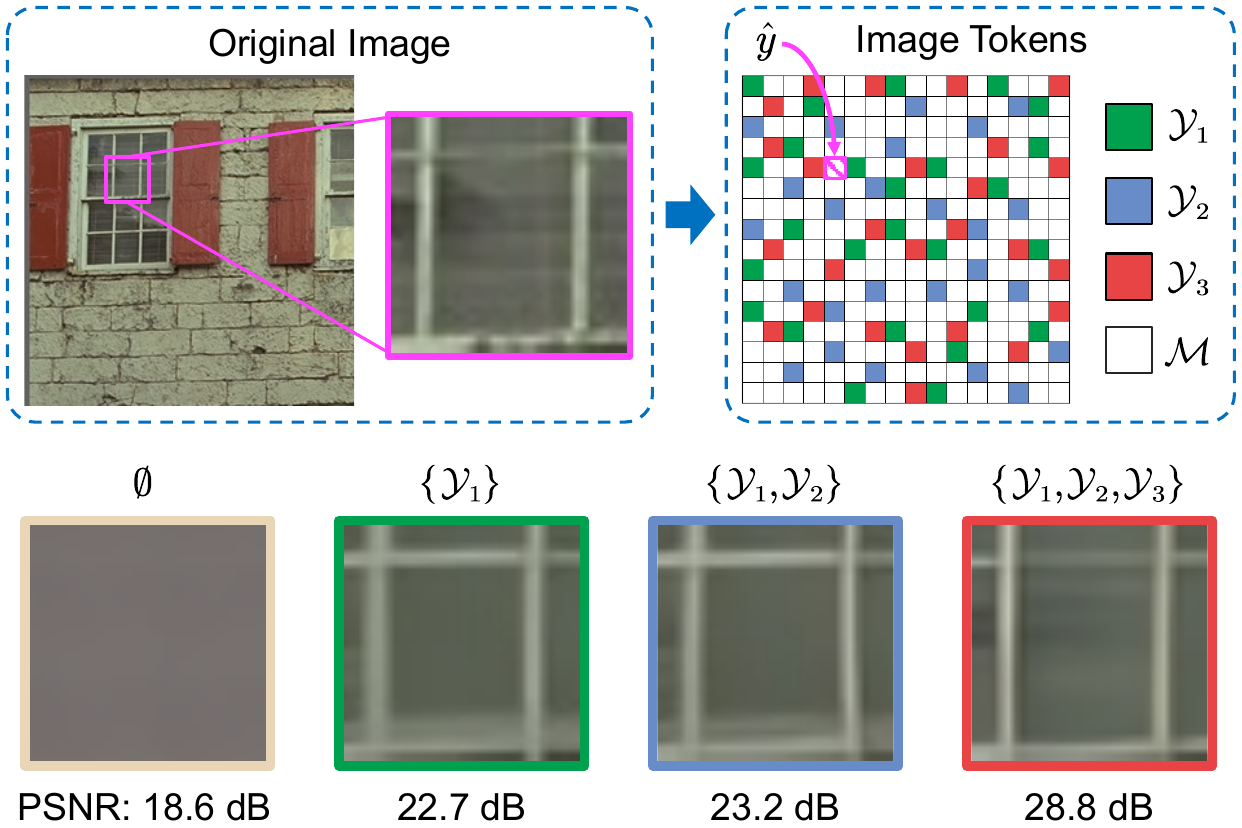}}
		\caption{Top left: original image (\emph{kodim01.png}) and a cropped patch. Top right: Three slices partitioned via QLDS location schedules with $L=10$ and $\beta=1.0$. It can be observed that all points in a slice are far from each other. We conduct latent PLC from partial token slices (missing values are filled with mask tokens) and decode them \emph{using the same model}. Bottom (left to right): reconstruct patches decoded from all mask tokens, slice 1 only, slice 1 and 2, slice 1 to 3, respectively.}
		\label{Fig_mask_token_prediction_a}
	\end{figure}

    \subsubsection{Function 1, for entropy modeling}
    The density modeling process can be expressed as:
    \begin{equation}
        \label{entropy_model}
        \bm{\varTheta} = h_{\text{density}}(f(\hatym)),
    \end{equation}
    where $\bm{\varTheta}$ encapsulates tokens' conditional distribution parameters.
    {During training, the rate term is only calculated for the masked locations. Consequently, the efficiency loss is formulated as follows:
    \begin{equation}
        \label{rate_loss}
        \mathcal{L}_{E} = \mathbb{E}_{\bm{x} \sim p_{\bm{x}}}
        \left[\sum_{\forall i, \mathbf{M}_i=1}-\log_{2}p(\haty_i | \hatym)  + \lambda d\left(\bm{x}, g_s(\haty)\right)\right],
    \end{equation}
    where $i$ specifies the spatial masking locations.}

    {During inference, similar to MaskGIT~\cite{chang2022maskgit}, the density modeling function operates iteratively. It begins with all masked tokens and uncovers a subset of tokens at each step until all tokens are revealed, and the rate term corresponds to the bitrate required to compress the full $\haty$.}
    
{   
	Specifically, in \eqref{rate_loss}, each token's distribution is factorized over the $C$ channels as:
	\begin{equation}
		\label{factorized_conditional_distribution_1}
		p(\haty_i | \hatym) = \prod_{c=1}^{C} p(y_i^c | \hatym),
	\end{equation}
	where $c$ represents the channel location, and the scalar $y_i^c$ denotes the element in $c$-th channel of the $i$-th mask token.
	
    To obtain a flexible and accurate density estimation for $y_i^c$, we exploit the Gaussian mixture model (GMM) with $K=3$ mixtures~\cite{cheng2020learned}.
    The density head predicts $3K$ parameters for each token element $y_i^c$.
    Specifically, the first $K$ dimensions are passed through a softmax activation to obtain the mixture probabilities $\bm{w}_i^c \in [0, 1]$.
    The next $K$ dimensions use a softplus activation to generate the variance parameters $\bm{\sigma}_i^c$. The final $K$ dimensions represent the mean parameters $\bm{\mu}_i^c$.

    In general, for a token element $\hat{y}_i^c$ in the $c$-th channel of the $i$-th token, its conditional distribution is formulated~as~follows:
    \begin{equation}
        \label{factorized_conditional_distribution_2}
        p(\hat{y}_i^c | \hatym) = \underset{\text{Gaussian Mixture Model}}{\underbrace{\left(\sum_{k=1}^{K} {w}_{i, k}^{c} \mathcal{N}\left({\mu}_{i, k}^{c}, ({\sigma}_{i, k}^{c})^2 \right) \right)}}(y_i^c + u),
    \end{equation}
    where additive i.i.d.\ noise $u \sim \mathcal{U}\left(-\frac{1}{2},\frac{1}{2}\right)$ is used to simulate quantization for entropy model during training~\cite{balle2020nonlinear}.
    
    }

    \subsubsection{Function 2, for latent PLC}
    The latent PLC process can be described by the following equation:
    \begin{equation}
        \label{latent_PLC_model}
        \checky = h_{\text{PLC}}(f(\hatym)) \odot \mathbf{M} + \haty \odot (\mathbf{1} - \mathbf{M}),
    \end{equation}
	where the binary mask $\mathbf{M}$ is used to simulate the impact of packet loss on image tokens.

    In fact, the actual distribution of packet loss ratios and patterns can vary a lot from the uniformly sampled masking ratio and independent dropping approach.
    However, training with the loss patterns, sampling from a fitted packet loss model (e.g., Markov model~\cite{749301}) or a fixed masking ratio, will hurt its generalization to unseen network conditions.
    Therefore, in practice, we sample from a universal masking ratio $r \sim \mathcal{U}(0, 1)$ once for each iteration and use the same $\hatym$ to complete two tasks.
    We empirically show it generalizes well for various context modes and packet loss models.
    We also explore other variant implementations for latent PLC, including using the expected mean in~\eqref{factorized_conditional_distribution_2} as an likelihood estimation to fill the lost values.
    We carefully compared with these methods detailed in the ablation study.

    \section{Loss-Resilient Image Compression}
    \label{sec:loss-resilient-image-compression}
	
	\subsection{Overview}

	\begin{figure}[tbp]
		\setlength{\abovecaptionskip}{0.cm}
		\setlength{\belowcaptionskip}{-0.cm}
		\centering
		{\includegraphics[width=0.35\textwidth]{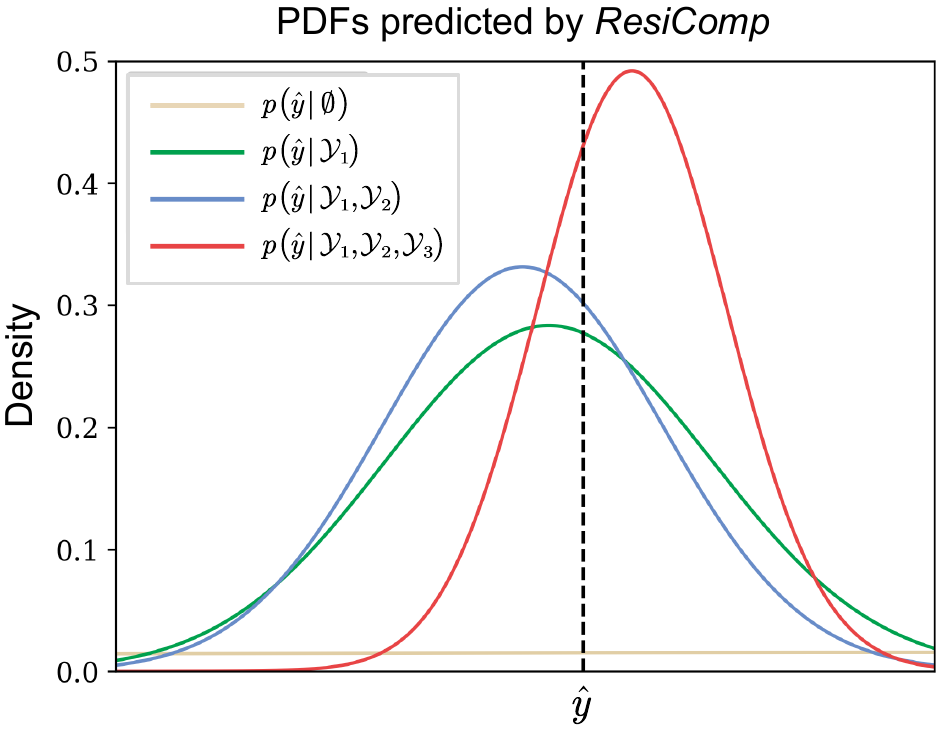}}
		\caption{Visualization of mask token prediction. It shows our Transformer enables flexible density predictions for $\hat{y}$, with accuracy improving as the number of context slices increases. The location of $\hat{y}$ is highlighted in Fig. \ref{Fig_mask_token_prediction_a}, and we visualize the channel with the highest entropy.}
		\label{Fig_mask_token_prediction_b}
	\end{figure}

    Given a dual-functional MVTM model trained as above, we now investigate various inference schemes to transmit~$\haty$ over a packet-lossy network.
    \emph{ResiComp} adopts a slice-based packetization and entropy coding scheme, with two advantages:
    \begin{enumerate}
        \item We limit the extend of error propagation across packets. When bit-errors occur in current packet, the receiver can re-synchronize at the next \emph{independently encoded packet}.
		\item The impact of packet loss can be considered as randomly replace token slices with all mask tokens, which aligns with the dual-functional MVTM training strategy introduced above.
    \end{enumerate}

    {Specifically, we divide the quantized tokens $\haty$ into multiple token slices $\mathcal{Y}_{l=1}^{L}=\{\mathcal{Y}_{1}, \dots, \mathcal{Y}_{L}\}$ using a partitioner shared between transmitter and receiver. 
    Each slice $\mathcal{Y}_{l}$ will be compressed into a bitstring using the entropy coding algorithm (e.g., arithmetic coding).
    According to information theory, conditional entropy is less than
    or equal to the entropy.
    To reduce the bitrate, some previously encoded slices serve as context $\mathcal{Y}_\text{ctx}$ in compressing current slice $\mathcal{Y}_{l}$.} 
    Intuitively, the more contexts we provided for encoding $\mathcal{Y}_{l}$, the better compression efficiency can be achieved, since $H(\mathcal{Y}_{l}) \geq H(\mathcal{Y}_{l} | \mathcal{Y}_{1}) \geq \cdots \geq H(\mathcal{Y}_{l} | \mathcal{Y}_{1}, \cdots, \mathcal{Y}_{l-1})$.
    
    However, in context of unreliable transmission, this approach can lead to severe error propagation.
    If any of its context packets $\{\mathcal{P}_{1}, \cdots, \mathcal{P}_{l-1}\}$ are missing, the slice $\mathcal{Y}_{l}$ fails to recover, leading to reduced resilience.
    {
    To mitigate this, existing error-resilient methods intentionally limit context utilization to enhance robustness~\cite{wang2000error}.
    }
    In \emph{ResiComp}, we follow this principle by using only a subset of pre-ordered slices as contexts, ensuring that errors in a single packet do not catastrophically hurt overall reconstruction quality. 

    	
    \begin{figure*}[ht]
    	\setlength{\abovecaptionskip}{0.cm}
    	\setlength{\belowcaptionskip}{-0.cm}
    	\centering
    	{\includegraphics[scale=0.42]{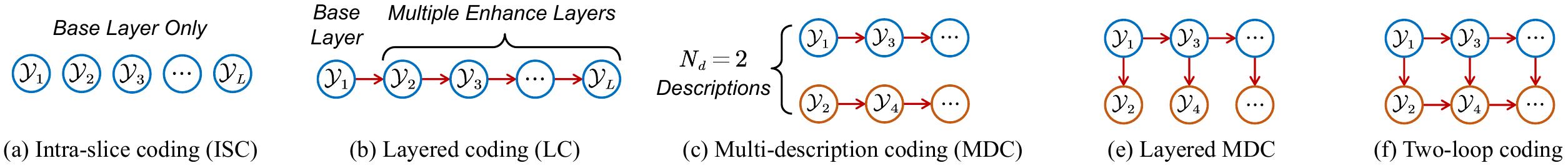}}
    	\caption{Our proposed context mode designs for trading off compression efficiency and resilience.}
    	\label{Fig_coding_mode}
    \end{figure*}
	
	\begin{figure}[t]
		\setlength{\abovecaptionskip}{0.cm}
		\setlength{\belowcaptionskip}{-0.cm}
		\centering
		{\includegraphics[width=0.45\textwidth]{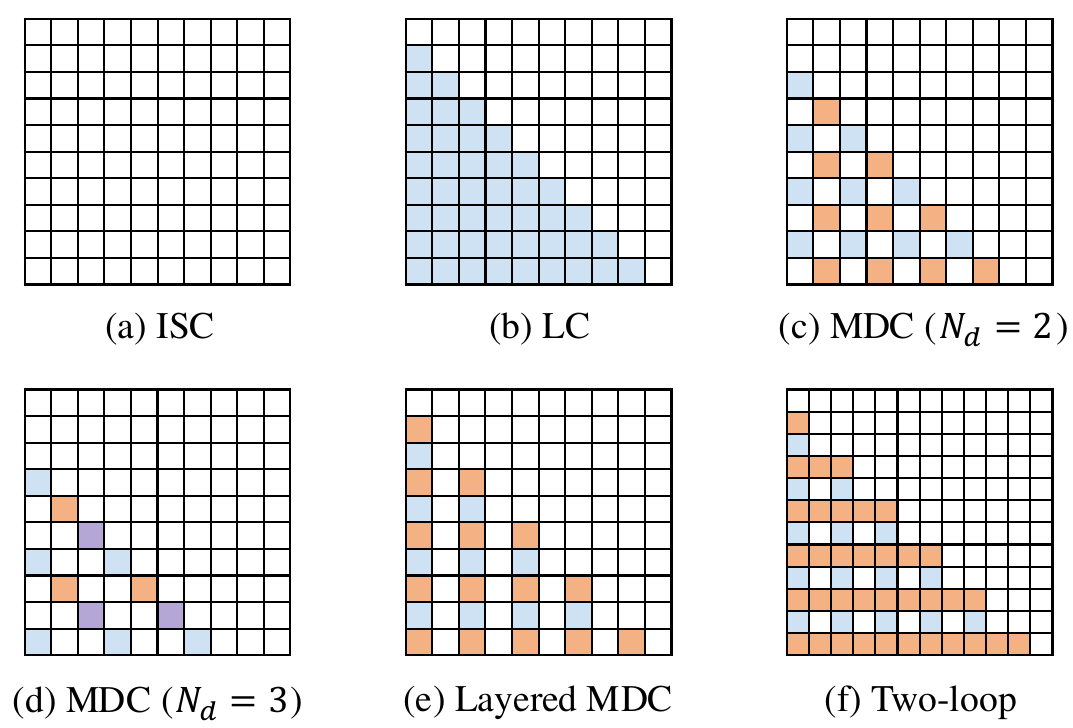}}
		\caption{(a)-(f) present the proposed context mode matrices, where a colored grid located in the $i$-th row and $j$-th column indicates that slice $\mathcal{Y}_i$ is predicted conditioning on $\mathcal{Y}_j$. Intuitively, coding schemes featuring more colored grids generally yield higher efficiency, whereas those with fewer colored grids tend to be more resilient. This leads to a crucial trade-off between compression efficiency and error resilience. }
		\label{Fig_context_matrix}
	\end{figure}

    \begin{algorithm}[t]
    	\caption{Dual-Functional Masked Visual Token Model}\label{algorithm_mvtm}
    	\textbf{Input:} Slice index $i$, context mode $\mathbf{G}$, packet received flag $\mathbf{F} \in \{0,1\}^{L}$ (only used at receiver), and mask token $\mathcal{M}$.
    	
    	\textbf{Output:} Predicted context vector $\mathcal{C}_i$
    	\begin{algorithmic}[1]
    		\Function{ContextTokenPrediction}{$i, \mathbf{G}, \mathbf{F}$}
    		\State $\mathcal{Y}_{\text{ctx}} \gets \emph{ones}(N, C) \cdot \mathcal{M}$\Comment{Initialize with all mask}
    		\For {$j \in \{1, \cdots, i-1\}$}
    		\State $m_j \gets  \text{partitioner.get\_locations(step=$j$)}$
    		\If{$\mathbf{G}[i,j] = 1$ and $\mathbf{F}_j = 1$}
    		\LineComment{Collect slice $j$ as contexts}
    		\State $\mathcal{Y}_{\text{ctx}}[m_j, :] \gets
    		\mathcal{Y}_{j}$
    		\ElsIf
    		{$\mathbf{G}[i,j] = 1$ and $\mathbf{F}_j = 0$}
    		\LineComment{Packet $i$ is lost due to error propagation}
    		\State \textbf{raise} \emph{SynchronizationException}
    		\EndIf
    		\EndFor
    		\LineComment{Context prediction with mask Transformer}
    		\State $\mathcal{C}_i \gets f(\mathcal{Y}_{\text{ctx}})[m_i, :]$
    		\State \Return context vector $\mathcal{C}_i$
    		\EndFunction
    	\end{algorithmic}
    \end{algorithm}

	\subsection{QLDS-Based Slice Partition} \label{subsec:slice-partition}

    The slice partition schedules in \emph{ResiComp} includes two important axes: (a) how many tokens are included in $\mathcal{Y}_{l}$, (b) which tokens are selected in $\mathcal{Y}_{l}$.
    To get precise prediction, we hope the mutual information between $\mathcal{Y}_{l}$ and \emph{all other groups} should be maximized.
    Intuitively, it not only benefits the context model by minimizing $H(\mathcal{Y}_{l} | \mathcal{Y}_\text{ctx})$, but also makes latent PLC easier at the same time.
    To achieve this, we build our slice partitioner based on the quantized low-discrepancy sequences (QLDS) schedule proposed in~\cite{mentzer2023m2t}.
    A QLDS in 2D is a sequence of points that pseudo-randomly traverse $N$ tokens, while minimizing the ``discrepancy'' for any given subsequence (refer to the Appendix of~\cite{mentzer2023m2t} for a detailed explanation).
    \emph{ResiComp} divides this QLDS into $L$ subsequences, with each specifying the specific tokens included in $\mathcal{Y}_{l}$.
    
    A visualization demo of QLDS-based slice partitioning are shown in Fig. \ref{Fig_mask_token_prediction_a}, where each image is encoded into a default~$L=10$~packets.
    In Fig. \ref{Fig_mask_token_prediction_b}, we further show a demo to illustrate how the mask token prediction benefits from contexts.
    To control the packet size distribution, we adapted the power schedule from MaskGIT~\cite{chang2022maskgit} to fit our flexible encoding structure.
    Specifically, the size of slice $l$ is determined as:
    \begin{equation}
        N_l = N \cdot \frac{(1 + C_l / L) ^ \beta}{\sum_{i=1}^{L} (1 + C_i / L) ^ \beta} ,
    \end{equation}
    where $C_l$ represents the number of context slices for $\mathcal{Y}_l$.
    The exponential factor $\beta$ is adjusted according to the specific encoding structure.

    \subsection{Slice-based Context Modeling}
    
\begin{algorithm}[t]
	\caption{ResiComp Sender}\label{algorithm_sender}
	\textbf{Input:} Input image $\bm{x}$, seed $s$, packet number $L$, mask token $\mathcal{M}$, and coding mode matrix $\mathbf{G}$
	
	\textbf{Output:} Set of packets $\{\mathcal{P}_1, \cdots, \mathcal{P}_L\}$
	\begin{algorithmic}[1]
		\Function{ResiComp Encoder}{$\bm{x}, L, \mathcal{M}, \mathbf{G}$}
		\State $\haty \gets Q\left(g_a(\bm{x})\right)$  \Comment{Tokenization}
		\State $\text{partitioner} \gets \text{qlds\_partitioner}(\text{seed=}s)$
		\State $\{\mathcal{Y}_1, \cdots, \mathcal{Y}_L\} \gets  \text{partitioner.get\_slices}(\haty, L)$
		\State $\mathbf{G} \gets \text{make\_context\_mode}(\text{coding\_mode}, L)$
		\For {$i \in \{1, \cdots, L\}$}
		\LineComment{Extract contexts of slice $i$}
		\State $\mathcal{C}_i \gets \Call{ContextTokenPrediction}{i, \mathbf{G}, \emph{ones}(L)})$
		\LineComment{GMM parameters estimation}
		\State $\bm{w}_i, \bm{\mu}_i, \bm{\sigma}_i \gets h_{\text{density}}\left(\mathcal{C}_i \right)$
		\LineComment{Slice-based entropy encoding \& packetization}
		\State $\mathcal{P}_i \gets \text{entropy\_encoder}(\mathcal{Y}_i, \bm{w}_i, \bm{\mu}_i, \bm{\sigma}_i)$
		\EndFor
		\State \Return packet sequence $\{\mathcal{P}_1, \cdots, \mathcal{P}_L\}$
		\EndFunction
	\end{algorithmic}
\end{algorithm}

\begin{figure}[tbp]
	\setlength{\abovecaptionskip}{0.cm}
	\setlength{\belowcaptionskip}{-0.cm}
	\centering
 	{\includegraphics[width=0.47\textwidth]{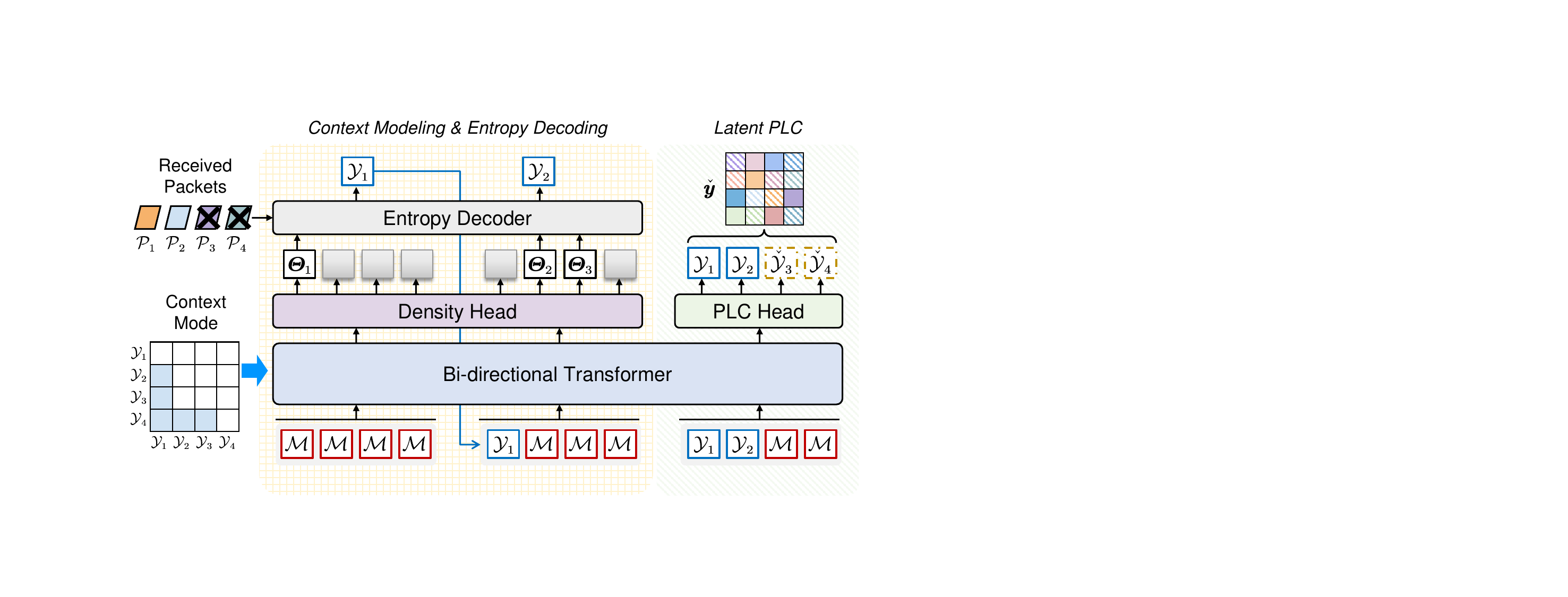}}
	\caption{Detailed process of context modeling, entropy decoding, and latent PLC. When packet loss occurs, \emph{ResiComp} first tries to decodes each received packet (some packets is received but undecodeable due to its context packets are lost), and then predicts the remaining tokens with latent PLC module.}
	\label{Fig_iterative_decoding}
\end{figure}

\begin{algorithm}[th]
	\caption{ResiComp Receiver}\label{algorithm_receiver}
	\textbf{Input:} Packet number $L$, seed $s$, coding structure $\mathbf{G}$, and packet is received flag~$\mathbf{F} \in \{0,1\}^{L}$
	
	\textbf{Output:} Reconstruction image $\checkx$
	\begin{algorithmic}[1]
		\Function{ResiComp Decoder}{$L, s, G, \mathbf{F}$}
		\State $\text{partitioner} \gets \text{make\_partitioner}( \text{partition\_mode}, \text{seed=}s)$
		\State $\hatym \gets \emph{ones}(N, C) \cdot \mathcal{M}$\Comment{Collect received tokens}
		\State $\mathbf{M} \gets \emph{ones}(N, C)$ \Comment{Mark locations to be decoded}
		\For {$i \in \{1, \cdots, L\}$}
		\Comment{De-packetization}
		\If{$\mathbf{F}_i = 1$}
		\LineComment{Try to collect contexts of slice $i$}
		\State $m_i \gets  \text{partitioner.get\_locations(step=i)}$
		\State \textbf{try:}
		\State \quad \quad {$\mathcal{Y}_{\text{ctx}} \gets$ \Call{CollectContexts}{$i, G, \mathbf{F}$}}
		\State \textbf{catch} \emph{SynchronizationException}:
		\State \quad \quad $\mathbf{F}_i \gets 0$
		\State \quad \quad \textbf{continue}
		
		\LineComment{All contexts received, recover slice $i$}
		\State $\bm{w}_i, \bm{\mu}_i, \bm{\sigma}_i \gets h_{\text{density}}\left(f(\mathcal{Y}_{\text{ctx}})\right)[m_i, :]$
		\State $\mathbf{M}[m_i, :] = 0$ \Comment{Mark decoded locations}
		\State $\hatym[m_i, :] \gets \text{entropy\_decoder}(\mathcal{P}_i, \bm{w}_i, \bm{\mu}_i, \bm{\sigma}_i)$
		\EndIf
		\EndFor
		\LineComment{Latent packet loss concealment}
		\State $\check{\bm{y}} \gets h_{\text{PLC}}\left(f(\hatym)\right) \cdot \mathbf{M} + \hatym \cdot (\bm{1} - \mathbf{M})$
		\State $\checkx \gets g_s(\check{\bm{y}})$
		\State \Return reconstruction image $\checkx$
		\EndFunction
	\end{algorithmic}
\end{algorithm}

    To explicitly control the efficiency-resilience trade-offs,    we propose to manage the contextual relationship between token slices using the idea of \emph{context mode}.

    A context mode is represented by a 0-1 valued matrix $\mathbf{G}~\in~\{0, 1\}^{L \times L}$, where $\mathbf{G}[l, k] = 1$ indicates that the $l$-th token slice $\mathcal{Y}_l$ is predicted conditioning on its preceding slice $\mathcal{Y}_k$.
    This matrix
    $\mathbf{G}$ exhibits the following properties:
    \begin{itemize}
        \item \textbf{Recoverability}:  $\mathbf{G}$ is a lower triangular matrix with all diagonal elements being zero.
        This structure ensures two critical aspects: (a) no slice $\mathcal{Y}_k$ is predicted based on itself, as indicated by the zero-valued diagonal.
        (b) each slice $\mathcal{Y}_k$ cannot have a subsequent slice $\mathcal{Y}_l$ (where $l>k$) as its context, preserving the unidirectional flow of information.
        Consequently, each packet is decodeable with its packet received and its contexts already decoded, enabling the decoder can reconstruct $\haty$ losslessly when all packets are successfully received.
        \item \textbf{Contextual Inheritance}: If $\mathcal{Y}_k$ acts as the context for $\mathcal{Y}_l$, then all contexts of $\mathcal{Y}_k$ also serve as contexts for $\mathcal{Y}_l$.
        This is because decoding $\mathcal{Y}_l$ requires the previous decoding of $\mathcal{Y}_k$, with the decoding of $\mathcal{Y}_k$ dependent on all its contexts being successfully decoded first.
        This setup allows for improved compression efficiency without sacrificing resilience.
    \end{itemize}

    \emph{ResiComp} facilitates flexible, customizable spatial-wise context modes.
    It is capable of integrating with any spatial context entropy model, such as the checkerboard~\cite{he2021checkerboard} or quincunx~\cite{el2022pqmim} patterns.
    This integration is achieved by modifying the slice partitioning approach and adjusting the context mode.
    We propose a range of modes, whose operational diagrams are illustrated in Fig. \ref{Fig_coding_mode}, with right arrows indicating the context order.
    The context mode matrices corresponding to these modes are depicted in Fig. \ref{Fig_context_matrix}.
    {Fig. \ref{Fig_iterative_decoding} presents an example to demonstrate our decoding progress in presence of packet loss, including iterative decoding and latent PLC.      
    The token slice decoding process requires no more than $L$ Transformer iterations, as slices sharing the same context can be predicted within the same forward pass step.
    Additionally, the GMM parameters of the independent token slice (e.g., $\mathcal{Y}_1$) are also cacheable, as they remain constant after training.

    With different context modes, \emph{ResiComp} can traverse multiple efficiency-resilience trade-offs within a single model adapted to network conditions, \emph{without additional training requirement}.}
    We summarize the masked token prediction process of our dual-functional MVTM in Algorithm~\ref{algorithm_mvtm}.
    The pseudo-code for the workflows of \emph{ResiComp} sender and receiver are shown in Algorithm~\ref{algorithm_sender} and Algorithm~\ref{algorithm_receiver}, respectively.

    \section{Experiments}\label{section_experiments}
    
    \begin{figure*}[ht]
	\setlength{\abovecaptionskip}{0.cm}
	\setlength{\belowcaptionskip}{-0.cm}
	\begin{minipage}[b]{0.31\textwidth}
		\centering
		\subfigure[EP1, $\epsilon=0.002, \gamma=6.50$]{\includegraphics[width=\textwidth,height=0.18\textheight]{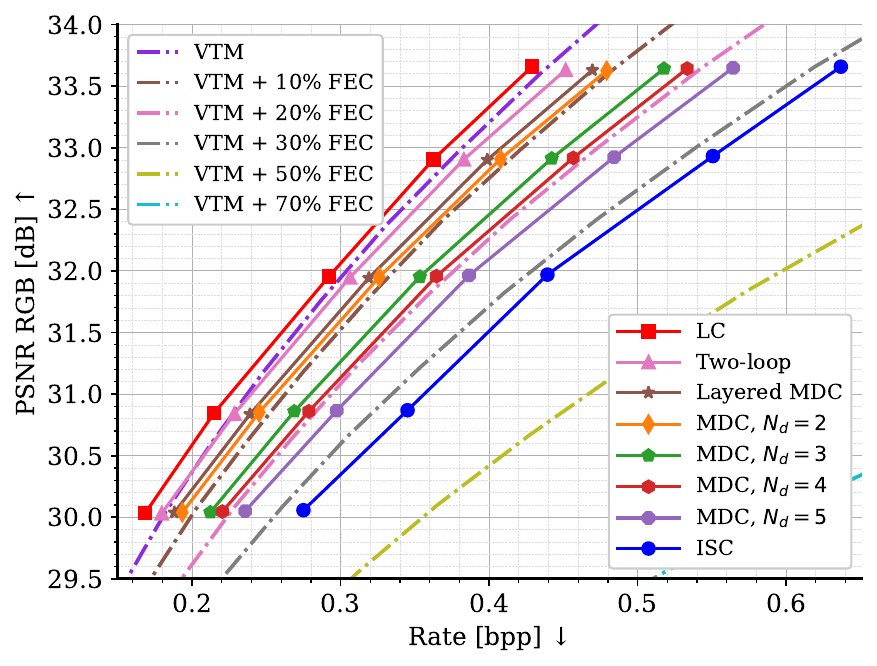}}
	\end{minipage}
	\quad
	\begin{minipage}[b]{0.31\textwidth}
		\centering
		\subfigure[EP2, $\epsilon=0.031, \gamma=1.59$]{\includegraphics[width=\textwidth,height=0.18\textheight]{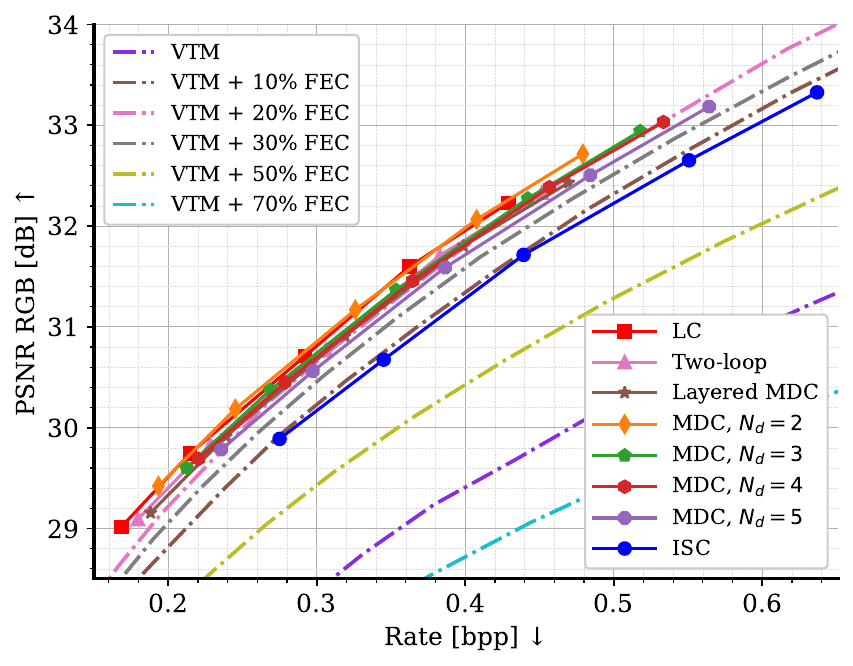}}
	\end{minipage}
	\quad
	\begin{minipage}[b]{0.31\textwidth}
		\centering
		\subfigure[EP3, $\epsilon=0.065, \gamma=5.00$]{\includegraphics[width=\textwidth,height=0.18\textheight]{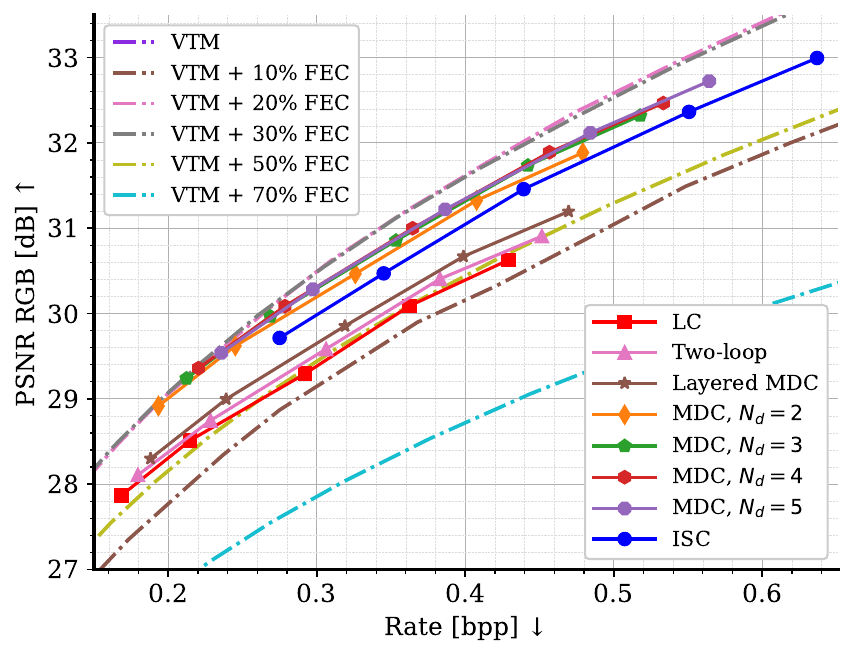}}
	\end{minipage}
	\\
	\begin{minipage}[b]{0.31\textwidth}
		\centering
		\subfigure[EP4, $\epsilon=0.136, \gamma=1.69$]{\includegraphics[width=\textwidth,height=0.18\textheight]{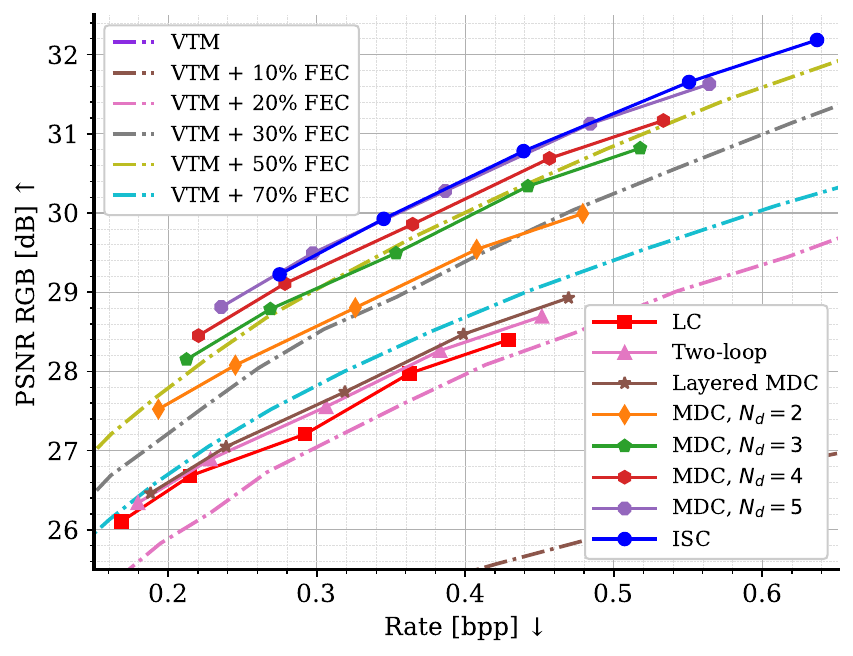}}
	\end{minipage}
	\quad
	\begin{minipage}[b]{0.31\textwidth}
		\centering
		\subfigure[EP5, $\epsilon=0.214, \gamma=10.0$]{\includegraphics[width=\textwidth,height=0.18\textheight]{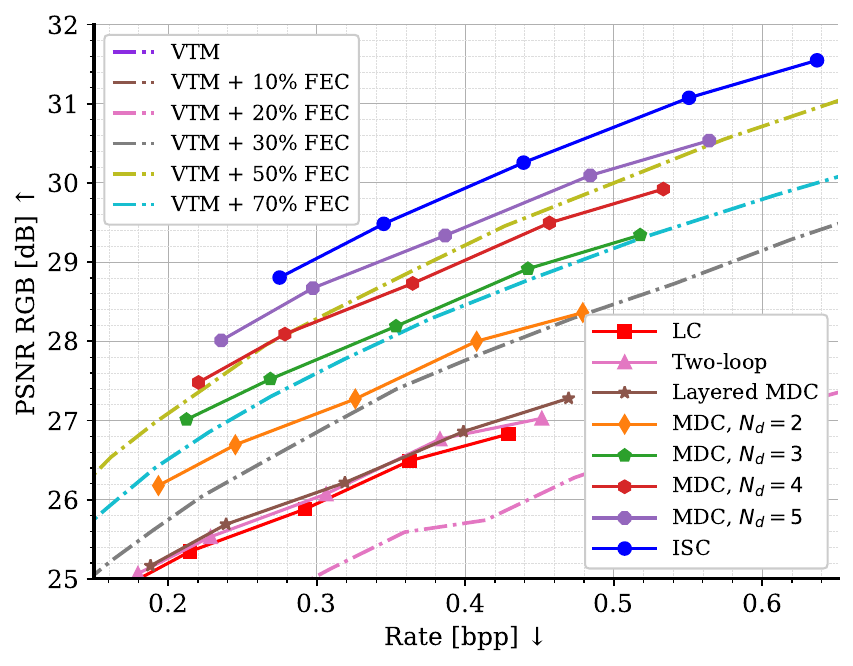}}
	\end{minipage}
	\quad
	\begin{minipage}[b]{0.31\textwidth}
		\centering
		\subfigure[EP6, $\epsilon=0.323, \gamma=2.71$]{\includegraphics[width=\textwidth,height=0.18\textheight]{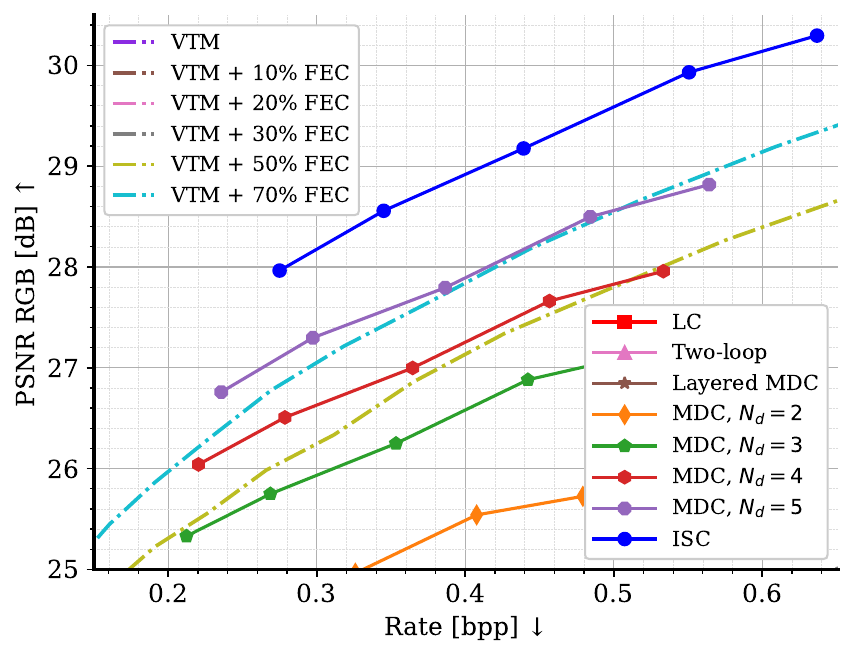}}
	\end{minipage}
	\caption{\emph{Packet-lossy transmission case:} RD performance comparison between our \emph{ResiComp} with diverse context modes (solid line, legends in right-bottom) and VTM + FEC (dotted dashed line, legends in left-top) over different packet lossy networks on the Kodak dataset.}
	\label{Fig_lossy_RD}
	\end{figure*}

    \subsection{Experimental Setup}\label{subsec:experimental-setup}

    \subsubsection{Training details}
    Our model is trained the DIV2K dataset~\cite{agustsson2017ntire}.
    During training, images are randomly cropped into $384 \times 384$ patches.
    We use the Adam optimizer with default parameters, and schedule the learning rate from $10^{-4}$ in initial 1M steps to $10^{-5}$ for another 100k steps.
    We follow the ``$\lambda$ warmup'' trick where $\lambda$ is set 10$\times$ higher during the first 15\% of training~\cite{mentzer2023m2t}.
    We first trained a MSE optimized model with $\lambda = 0.0035, \alpha=0.1$, and then finetuned it to other variants $\lambda \in \{0.0017, 0.0025, 0.005, 0.007\}$.

    \subsubsection{Evaluation datasets}
    Our evaluation mainly uses the widely used Kodak~\cite{Kodak} with 24 uncompressed images at the size of $512 \times 768$.
    We also consider the CLIC Professional Valid~\cite{CLIC21} with 30 images to validate the robustness on higher resolution.
    For neural codecs, we pad the image to multiples of 16 pixels during preprocessing, and then unpad the reconstruction to its origin resolution.

    \subsubsection{Evaluation configurations}
    To evaluate compression efficiency, we compare the RD curves under the assumption of reliable transmission. The comparison includes advanced traditional codecs, such as BPG~\cite{BPG}, VTM-20.0~\cite{VVC_intra}, and neural codecs, namely Hyperprior~\cite{balle2018}, JointAR~\cite{minnen2018}, GMM-Attn~\cite{cheng2020learned}, ELIC~\cite{he2022elic}, and very recent MLIC \cite{jiang2023mlic} and TCM~\cite{liu2023learned}.

    To evaluate the loss resilience capability, we use the state-of-the-art traditional codec, VTM-20.0~\cite{VVC_intra}, combined with an ideal FEC ($\rho=0$) \cite{4427233} for protection, denoted as ``VTM + $\text{x}\%$ FEC''.
    In this context, {the parity packets ratio $\text{x}\%$ is defined as $N_r/(N_k + N_r)$.
    For instance, the VTM + $30\%$ FEC scheme encodes an image into $N_k = 7$ data packets and $N_r = 3$ parity packets. 
    The image can be decoded successfully at the receiver if \emph{any $7$ packets} out of total $10$ packets are received. 
    However, this requires additional bandwidth to transmit the parity packets.
    In this example, the bandwidth cost of VTM + $30\%$ FEC will be $(N_k + N_r)/N_k \approx 1.43$x compared to using VTM alone.
    }
    For a comprehensive comparison, we also consider the widely implemented layered coding with unequal error protection (LC + UEP) technique for comparison, and provide backup packets to protect their base layer.
    Metrics used in our evaluation included bit per pixel (bpp) for rate measurement and Peak Signal-to-Noise Ratio (PSNR) for quality assessment.
    In scenarios where the receiver cannot decode the transmitted image, we assigned a consistent PSNR of 13 dB to all schemes.

    \subsubsection{Packet loss models}
    
    	\begin{figure}[t]
    	\setlength{\abovecaptionskip}{0.cm}
    	\setlength{\belowcaptionskip}{-0.cm}
    	\centering
    	{\includegraphics[width=0.35\textwidth]{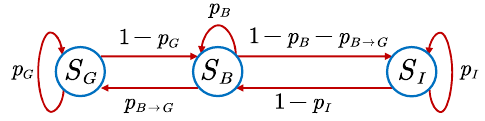}}
    	\caption{The 3-state Markov model used for packet-level simulation.}
    	\label{Fig_packet_loss_model}
    \end{figure}
    
    \begin{table}[t]
    	\renewcommand{\arraystretch}{1.3}
    	\centering
    	\caption{Three-state Markov chain parameters}
    	\small
    	\begin{tabular}{m{0.7cm}<{\raggedright}|c c c c c c}
    		\Xhline{1pt}
    		\centering{Case} & $p_G$ & $p_B$ & $p_I$ & $p_{B \rightarrow G}$ & $\epsilon$ & $\gamma$ \tabularnewline
    		\Xhline{0.6pt}
    		\centering{EP1} & 0.99968 & 0.8462 & 0.0000 & 0.1538 & 0.002 & 6.50 \tabularnewline
    		\centering{EP2} & 0.9798 & 0.3720 & 0.3333 & 0.6304 & 0.031 & 1.59 \tabularnewline
    		\centering{EP3} & 0.9500 & 0.8000 & 0.6000 & 0.8000 & 0.065 & 5.00 \tabularnewline
    		\centering{EP4} & 0.9363 & 0.4072 & 0.5662 & 0.3631 & 0.138 & 1.69 \tabularnewline
    		\centering{EP5} & 0.9000 & 0.9000 & 0.1000 & 0.1000 & 0.214 & 10.0 \tabularnewline
    		\centering{EP6} & 0.8507 & 0.6305 & 0.2000 & 0.2982 & 0.323 & 2.71 \tabularnewline
    		\Xhline{1pt}
    	\end{tabular}
    	\label{Table_packet_loss_model_params}
    \end{table}

	\begin{figure*}[t]
		\setlength{\abovecaptionskip}{0.cm}
		\setlength{\belowcaptionskip}{-0.cm}
		\centering
		\begin{minipage}{.19\textwidth}
			\centering
			\subfigure[LC, {\color{red}$23.9\%$}]{\includegraphics[width=\textwidth, height=0.13\textheight]{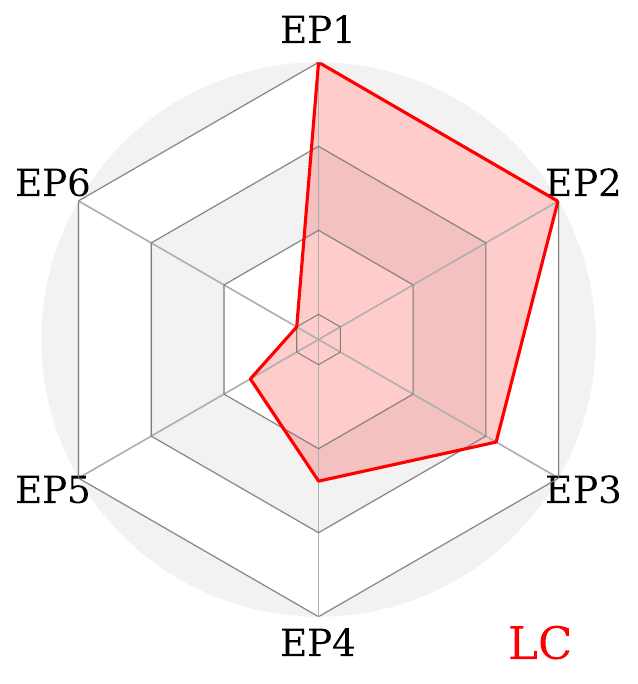}}
		\end{minipage}
		\begin{minipage}{.19\textwidth}
			\centering
			\subfigure[MDC ($N_d=2$), {\color{red}$14.1\%$}]{\includegraphics[width=\textwidth, height=0.13\textheight]{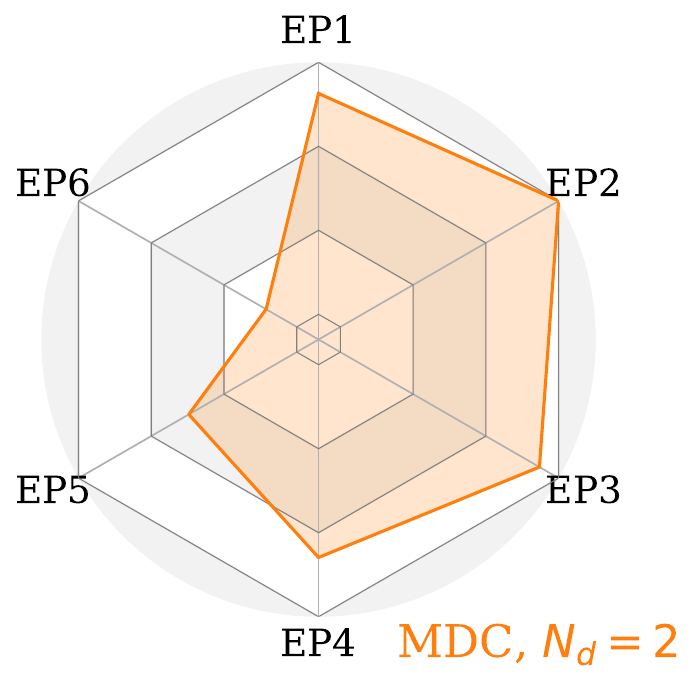}}
		\end{minipage}
		\begin{minipage}{.19\textwidth}
			\centering
			\subfigure[MDC ($N_d=4$), {\color{red}$6.5\%$}]{\includegraphics[width=\textwidth, height=0.13\textheight]{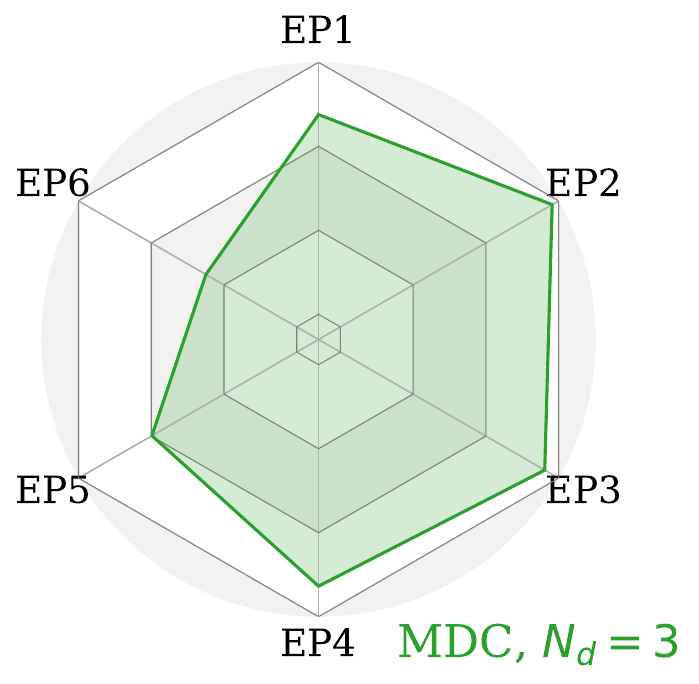}}
		\end{minipage}
		\begin{minipage}{.19\textwidth}
			\centering
			\subfigure[MDC ($N_d=5$), {\color{blue}$-2.7\%$}]{\includegraphics[width=\textwidth, height=0.13\textheight]{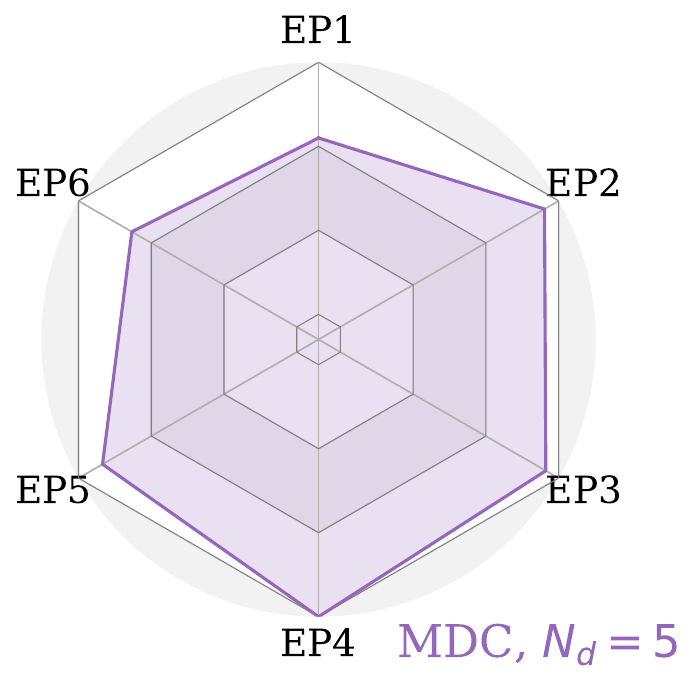}}
		\end{minipage}
		\begin{minipage}{.19\textwidth}
			\centering
			\subfigure[ISC, {\color{blue}$-17.7\%$}]{\includegraphics[width=\textwidth, height=0.13\textheight]{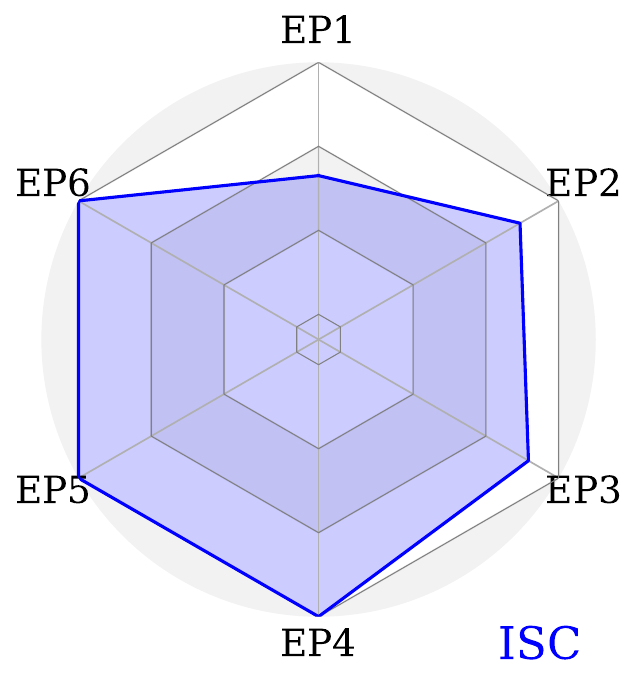}}
		\end{minipage}
		\\
		\begin{minipage}{.19\textwidth}
			\centering
			\subfigure[VTM, {\color{red}$21.5\%$}]{\includegraphics[width=\textwidth, height=0.13\textheight]{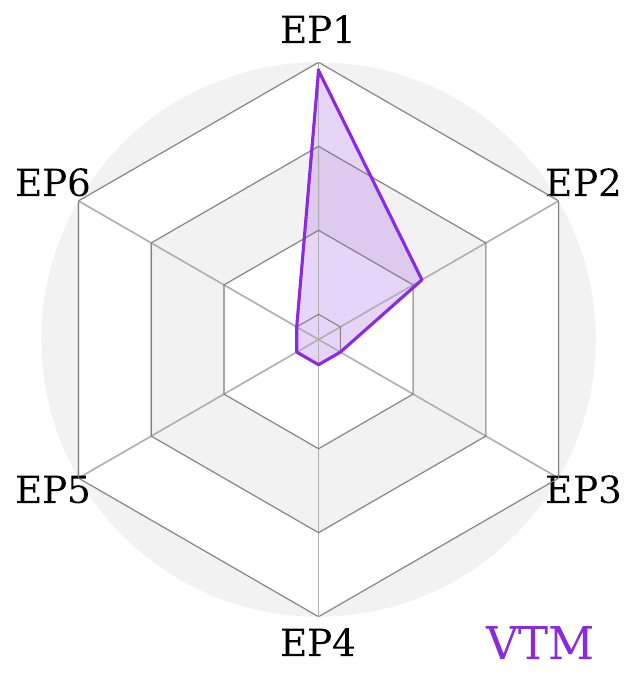}}
		\end{minipage}
		\begin{minipage}{.19\textwidth}
			\centering
			\subfigure[+ $10\%$ FEC, {\color{red}$12.8\%$}]{\includegraphics[width=\textwidth, height=0.13\textheight]{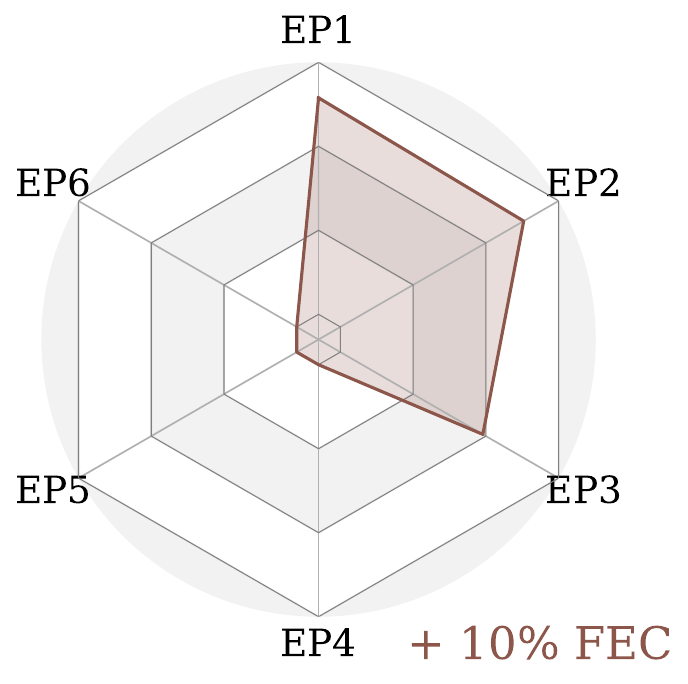}}
		\end{minipage}
		\begin{minipage}{.19\textwidth}
			\centering
			\subfigure[+ $30\%$ FEC, {\color{blue}$-12.2\%$}]{\includegraphics[width=\textwidth, height=0.13\textheight]{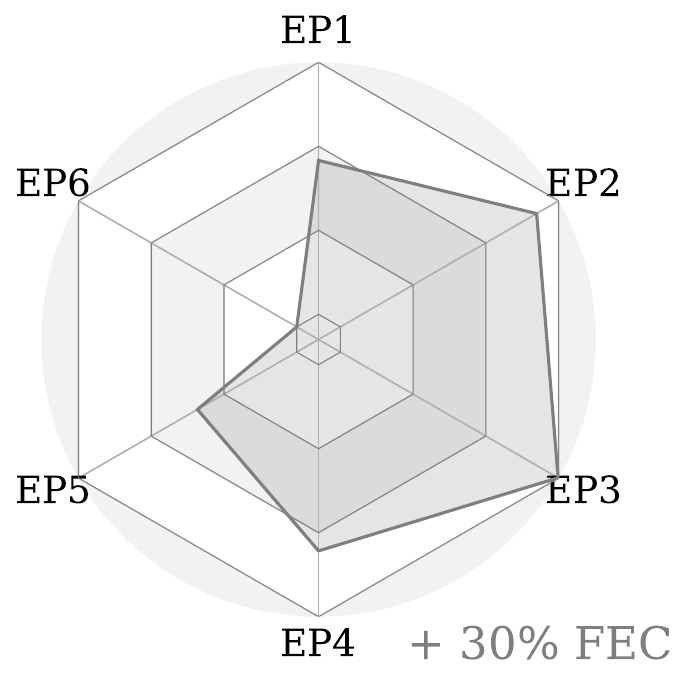}}
		\end{minipage}
		\begin{minipage}{.19\textwidth}
			\centering
			\subfigure[+ $50\%$ FEC,~{\color{blue}$-57.0\%$}]{\includegraphics[width=\textwidth, height=0.13\textheight]{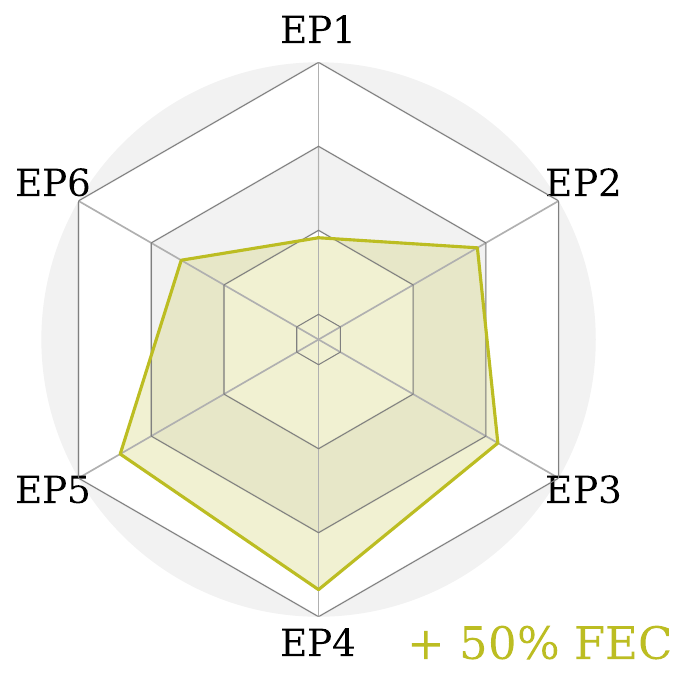}}
		\end{minipage}
		\begin{minipage}{.19\textwidth}
			\centering
			\subfigure[+ $70\%$ FEC,~{\color{blue}$-161.7\%$}]{\includegraphics[width=\textwidth, height=0.13\textheight]{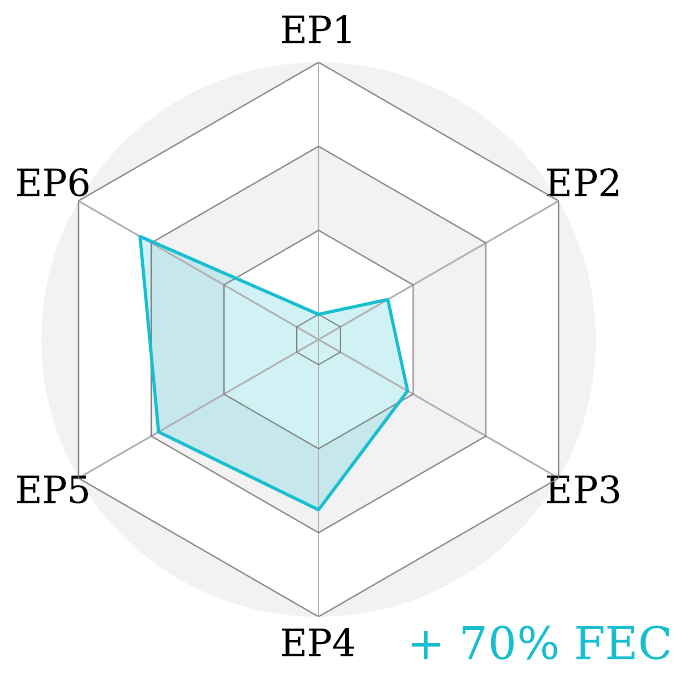}}
		\end{minipage}
		\caption{\emph{Packet-lossy transmission case:} Average PSNR across different packet lossy networks on the Kodak dataset.
			We indicate the bitrate savings compared to BPG in the figure captions, where {\color{red}red} denotes bitrate savings and {\color{blue}blue} represents extra bitrate costs. When comparing column by column, \emph{ResiComp} exhibits better resilience  (evidenced by a larger coverage area) compared to VTM + FEC at a lower bandwidth cost.}
		\label{Fig_radar}
	\end{figure*}

    In comparing loss-resilient schemes, we conduct packet-level simulations over diverse WLAN or mobile network conditions.
    In specific, to reproduce their error patterns, we utilized the packet loss traces simulated by the 3-state Markov model.
    The Markov chain-based models~\cite{749301} are widely-used and demonstrated effective in reproducing the characteristics of real network traces, especially for burst-like conditions with consecutive packet loss.
    Fig. \ref{Fig_packet_loss_model} illustrates the 3-state Markov chain used in our study, comprising a non-loss state (Good state, $S_G$), a lossy state (Bad state, $S_B$), and an intermediary state ($S_I$).
    Only the bad state $S_B$ presents packet loss, whereas the other two states are loss-free states~\cite{milner2004analysis}.

    To provide a thorough evaluation, we test \emph{ResiComp} and other resilient methods on $6$ error patterns (EP1 to EP6), with Markov chain parameters derived from real mobile network traces (EP1, EP2, and EP6), and from WLAN packet loss data (EP4).
    Key parameters of these packet loss models are detailed in Table~\ref{Table_packet_loss_model_params}, where $\epsilon$ indicates the probability of a packet being lost, and $\gamma$ denotes the average packet loss burst length.
    
    \subsection{Efficiency and Resilience Comparison in Lossy Networks} \label{resilience-comparison}
    
	We perform packet-level simulations using a 3-state Markov model trained on real WiFi and mobile network traces.
	Fig. \ref{Fig_lossy_RD} show the RD curves under packet-lossy network with various packet loss conditions (EP1 to EP6). 
	Here, we compare the resilience of \emph{ResiComp} ($\alpha=0.1$, with legends in the bottom-right) using different context modes, against VTM + FEC (with legends in the top-left).
	
	From the results, it is clear that by switching context modes from LC to ISC, our \emph{ResiComp} achieves a gradual shift from prioritizing high efficiency with promising resilience to focusing on high resilience with promising efficiency.
    When comparing the best performance curves of VTM + FEC and \emph{ResiComp} in each scenario, it is evident that \emph{ResiComp} surpasses or matches the performance VTM + FEC in all packet loss rate scenarios.
    Moreover, the performance gap between \emph{ResiComp} and VTM + FEC increases by a large margin in high packet loss rate scenarios (EP4 to EP6).
    The reason is that, as the average packet loss rate and burst length increase, there is decreased probability for parity packets' number perfectly aligning with the actual number of packets lost.
    Consequently, VTM + FEC requires a significantly higher parity packet ratio than average packet loss rate $\epsilon$, such as adding $50\%$ redundancy for EP4 ($\epsilon=0.136$) and $70\%$ redundancy for EP6 ($\epsilon=0.323$).
    This excessive redundancy hinders the effectiveness of FEC codes.

    To provide a more intuitively comparison for resilient transmission case, we compare mean PSNR values across $6$ scenarios using radar charts in Fig. \ref{Fig_radar}.
    The mean PSNR values is calculated over a bitrate integral range of $[0.3, 0.4]$ bpp to reflect the resilience across all codecs, and each scenario's results is normalized by dividing its maximum value.
    It is clear that while each VTM + FEC scheme excels in several EPs, it falls short in others.
    Their performance is highly sensitive to variations in packet loss rates and burst lengths, which poses challenges to transmitter in setting and adjusting the FEC redundancy rate.
    In contrast, \emph{ResiComp} exhibits more consistent resilience across all error patterns, covering a larger area in the radar charts than the VTM + FEC schemes.

        \begin{figure}[t]
    	\setlength{\abovecaptionskip}{0.cm}
    	\setlength{\belowcaptionskip}{-0.cm}
    	\centering
    	{\includegraphics[width=0.45\textwidth]{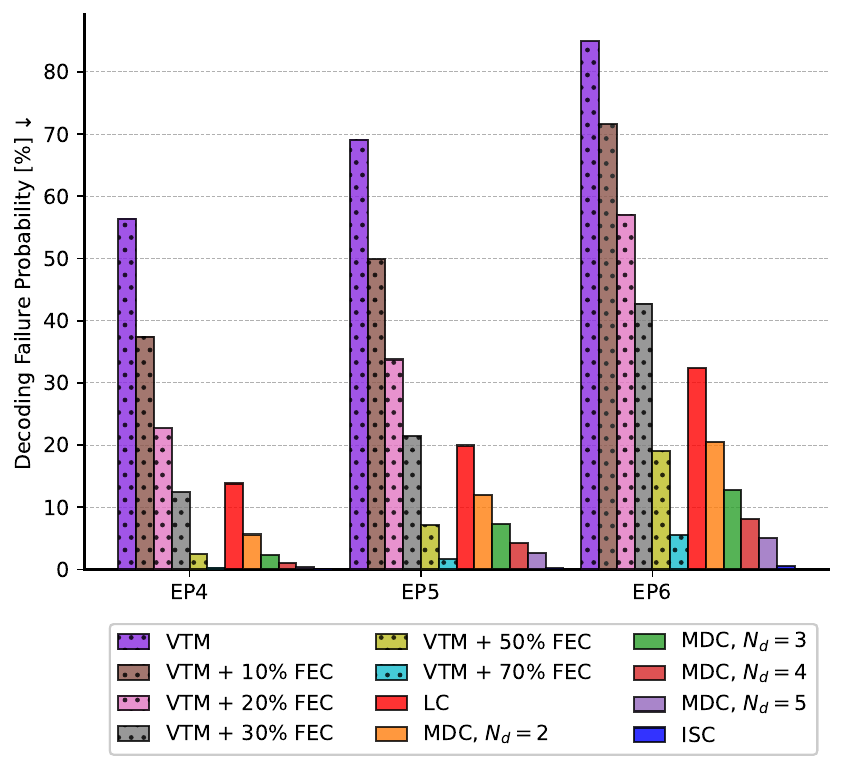}}
    	\caption{\emph{Packet-lossy transmission case:} the decoding failure probability during packet-level simulation. 
    		For VTM + $\text{x}\%$ FEC, decoding process is failed if less than $N_k$ out of total $N_k + N_r$ packets are received. 
    		For \emph{ResiComp}, decoding process is failed if no slice can be entropy decoded, i.e., $\hatym$ is all mask tokens. 
    		We select some representative resilient schemes and scenarios for clear explanation.}
    	\label{Fig_failed_prob}
    \end{figure}
    
    To further illustrate this point, the decoding failure ratios are presented in Fig. \ref{Fig_failed_prob}.
    These results indicate that, when achieving similar compression efficiency, \emph{ResiComp} maintains a much lower probability to break down than VTM + FEC\@.
    This observation verifies our choice to adapt the codec structure via context modes instead of adding redundancy, which is more robust against changing packet loss model characteristics.

    Figure \ref{Fig_benchmark} in Section \ref{section_introduction} summarizes the reliable and resilient transmission case within a efficiency-resilience benchmark.
    In specific, we use the BD-rate metric to assess bitrate savings (with BPG as an anchor), representing efficiency, and we calculate the mean PSNR averaged across $6$ lossy scenarios as the resilience metric.
    The results indicate that our ResiComp not only significantly enhances the resilience of NICs but also achieves a more favorable efficiency-resilience balance compared to VTM + FEC\@.
    Moreover, this trade-off is effectively adjustable by scheduling the spatial dependencies of tokens within a single model, with the LC mode offers the top tier efficiency, and the ISC mode excels in resilience.

    \subsection{Efficiency Comparison in Reliable Networks} \label{efficency-comparison}
    
    	\makeatletter
    \renewcommand{\@thesubfigure}{\hskip\subfiglabelskip}
    \makeatother
    
    \begin{figure}[t]
    	\setlength{\abovecaptionskip}{0.cm}
    	\setlength{\belowcaptionskip}{-0.cm}
    	\centering
    	\begin{minipage}{0.24\textwidth}
    		\centering
    		\subfigure[]{\includegraphics[width=\textwidth]{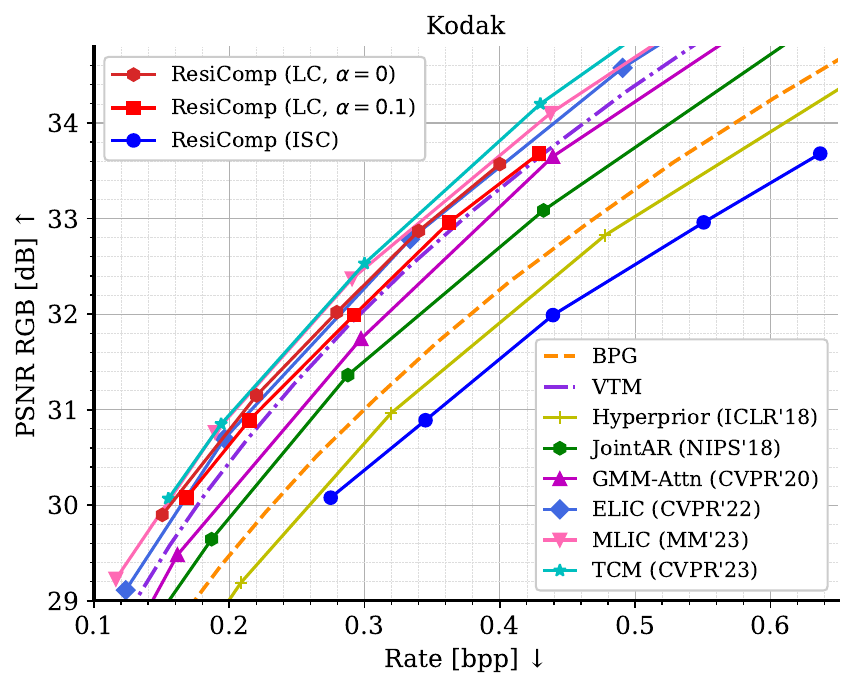}}
    	\end{minipage}
    	\begin{minipage}{0.24\textwidth}
    		\centering
    		\subfigure[]{\includegraphics[width=\textwidth]{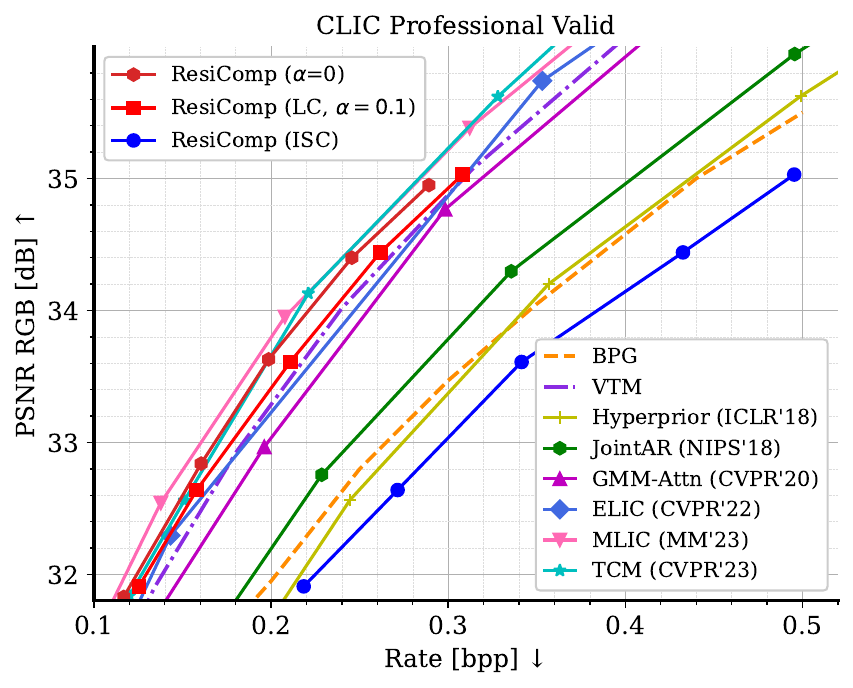}}
    	\end{minipage}
    	\vspace{-1.5em}
    	\caption{{\emph{Reliable transmission case}: RD performance comparison in a packet-loss-free network on Kodak and CLIC professional valid dataset.}}
    	\label{Fig_errorfree_RD}
    \end{figure}

	The rate-distortion performance on the Kodak dataset is shown in Fig. \ref{Fig_errorfree_RD}. The results for Hyperprior~\cite{balle2018}, JointAR~\cite{minnen2018}, and GMM-Attn~\cite{cheng2020learned} are obtained from the implementations of CompressAI~\cite{begaint2020compressai}, while the results for more recent NICs, including ELIC~\cite{he2022elic}, MLIC~\cite{jiang2023mlic}, and TCM~\cite{liu2023learned}, are taken from their respective open-source implementations.
	We observe that our \emph{ResiComp}, employing a layered coding mode (LC) with $\alpha=0.1$, outperforms the state-of-the-art traditional codec VTM. Compared to leading NICs like MLIC and TCM, \emph{ResiComp} incurs only a small reduction in RD performance while achieving significant improvements in resilience.
	As for \emph{ResiComp} (LC, $\alpha=0$), which focuses on RD optimization by removing resilience loss, it provides comparable RD performance to leading NICs, validating the effectiveness of our MVTM training with density head.
	Moreover, both MLIC and TCM use more expressive and complex nonlinear transforms than the ELIC transform in our approach, indicating potential for further performance improvement for our model.

    \subsection{ResiComp for Fine-Grained Progressive Decoding}

    	\begin{figure}[t]
    	\setlength{\abovecaptionskip}{0.cm}
    	\setlength{\belowcaptionskip}{-0.cm}
    	\centering
    	\begin{minipage}{0.24\textwidth}
    		\centering
    		\subfigure[]{\includegraphics[width=\textwidth]{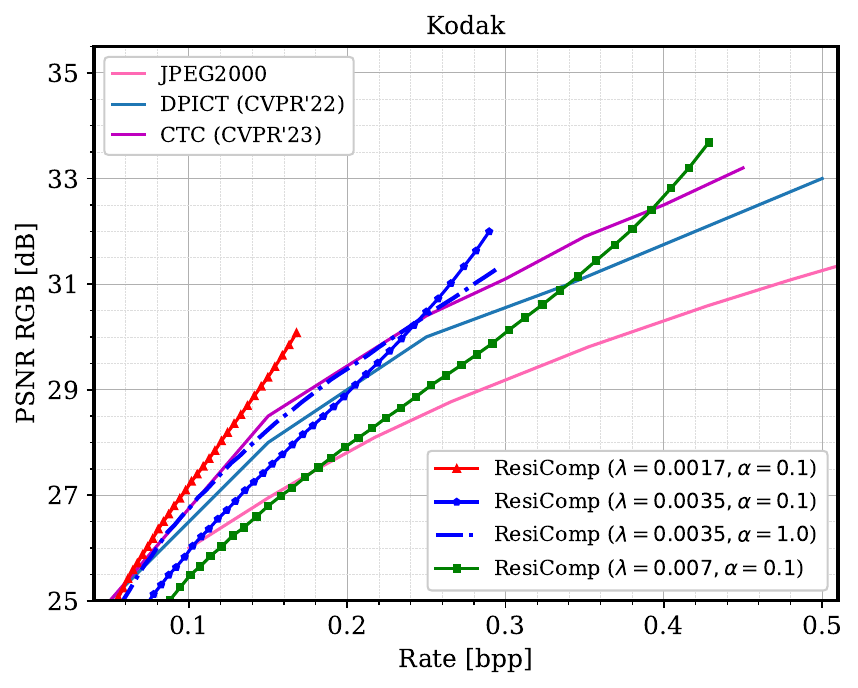}}
    	\end{minipage}
    	\begin{minipage}{0.24\textwidth}
    		\centering
    		\subfigure[]{\includegraphics[width=\textwidth]{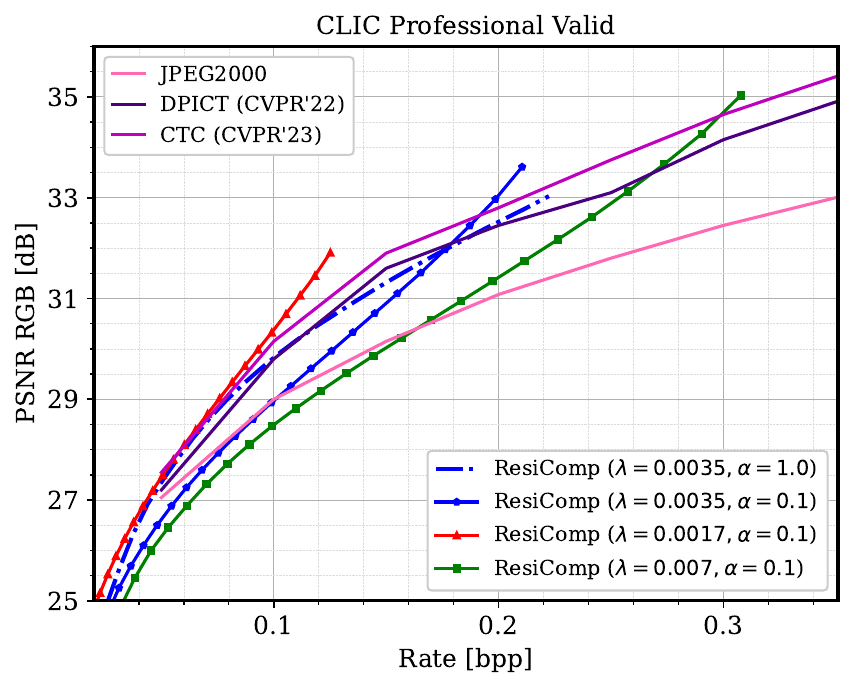}}
    	\end{minipage}
    	\vspace{-1.5em}
    	\caption{{\emph{Progressive coding case}: RD performance comparison on the (a) Kodak (b) CLIC professional valid dataset.}}
    	\label{Fig_progressive}
    \end{figure}
    
    The progressive compression aims to encode an image into a bitstream that can be truncate at multiple points to reconstruct at different qualities.
    It is feasible especially for the bandwidth-limited networks, a preview image can be quickly displayed first and then decoding the remaining bits for progressive quality improvement.
    From the perspective of resilient coding, this problem can be viewed as delivering input image over a special packet loss model, which always drops the tailed~packets.
    Thus, our proposed \emph{ResiComp} naturally supports for fine-grained progressive decoding, since the \emph{recoverability} property promises the tailed lost packets will not affect the decoding process in previous packets.

    In Fig. \ref{Fig_progressive}, we compare the RD curves of \emph{ResiComp} (LC) with recent advanced neural progressive codecs~\cite{ctc, dpict}. 
    Here, \emph{ResiComp} is configured with a larger slice number, $L=32$, where approximately 3\% unrecovered  tokens are decoded each time between two successive RD points.
    It can be observed that \emph{ResiComp} ($\alpha=0.1$) demonstrates superior efficiency when the majority of tokens are decoded. However, it exhibits a shorter rate region and underperforms compared to the trip-plane codecs~\cite{ctc, dpict} when only a portion of tokens are decoded. This is because \emph{ResiComp} is primarily trained for latent PLC in causal orders, leading to a trade-off in performance due to fixed context modeling orders. Despite this, we emphasize the potential of \emph{ResiComp}, particularly its dual-functional context modeling, as a promising fine-grained scalable coding method.

	\subsection{Ablation Study}\label{subsec:ablation-study}

		\begin{figure}[t]
		\setlength{\abovecaptionskip}{0.cm}
		\setlength{\belowcaptionskip}{-0.cm}
		\centering
		{\includegraphics[width=0.37\textwidth]{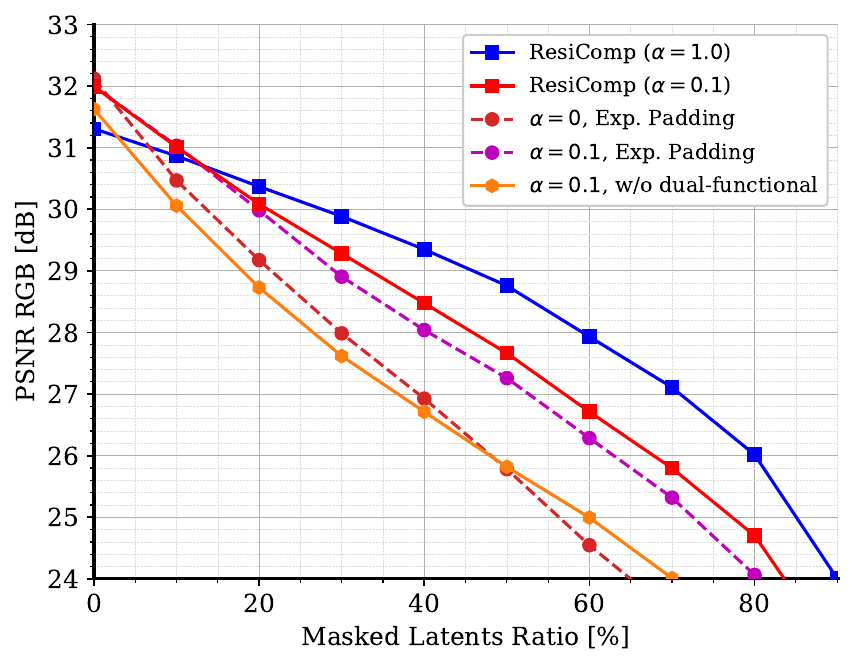}}
		\caption{Ablation studies on variant implementations for latent PLC on the Kodak dataset. We also vary the trade-off factor $\alpha$ to assess its impact. For a fair comparison, the bpp of all methods is set to approximately $0.3$. }
		\vspace{0em}
		\label{Fig_ablation_alpha}
	\end{figure}

    \subsubsection{Packet loss concealment alternatives}
    Fig.~\ref{Fig_ablation_alpha} illustrates our ablation studies on PLC alternatives, where we compare reconstruction quality with partially masked tokens.
    In the case of the two dashed curves, we removed the PLC head and filled the missing values with expected mean values from the predicted GMM distributions before decoding the latents.
    Our \emph{ResiComp} ($\alpha=0.1$) shows a notable improvement in resilience, particularly at larger masked ratios, while sacrificing minimal compression capability (approximately 0.2dB PSNR at at $0.3$ bpp).
    Additionally, increasing the value of $\alpha$ could further trade off some compression efficiency for even more substantial resilience improvements.
    Moreover, utilizing separate Transformers for the entropy model and latent PLC module (as indicated by the w/o dual-functional legend) results in a noticeable decline in both compression efficiency and loss resilience, which verifies the effectiveness of our proposed dual-functional approach.

    \subsubsection{Comparing with layered coding combined with UEP}

    \begin{figure}[t]
    	\setlength{\abovecaptionskip}{0.cm}
    	\setlength{\belowcaptionskip}{-0.cm}
    	\centering
    	{\includegraphics[scale=0.5]{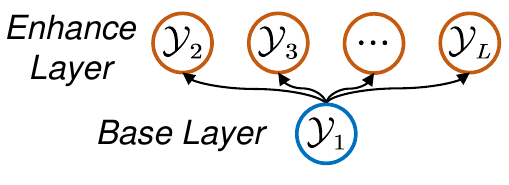}}
    	\caption{The operational diagram of scalable layered coding mode, incorporating one enhance layer.}
    	\label{Fig_scalable_layered_coding}
    	\vspace{0em}
    \end{figure}

    	\begin{figure}[t]
    	\setlength{\abovecaptionskip}{0.cm}
    	\setlength{\belowcaptionskip}{-0.cm}
    	\centering
    	\includegraphics[width=0.4\textwidth]{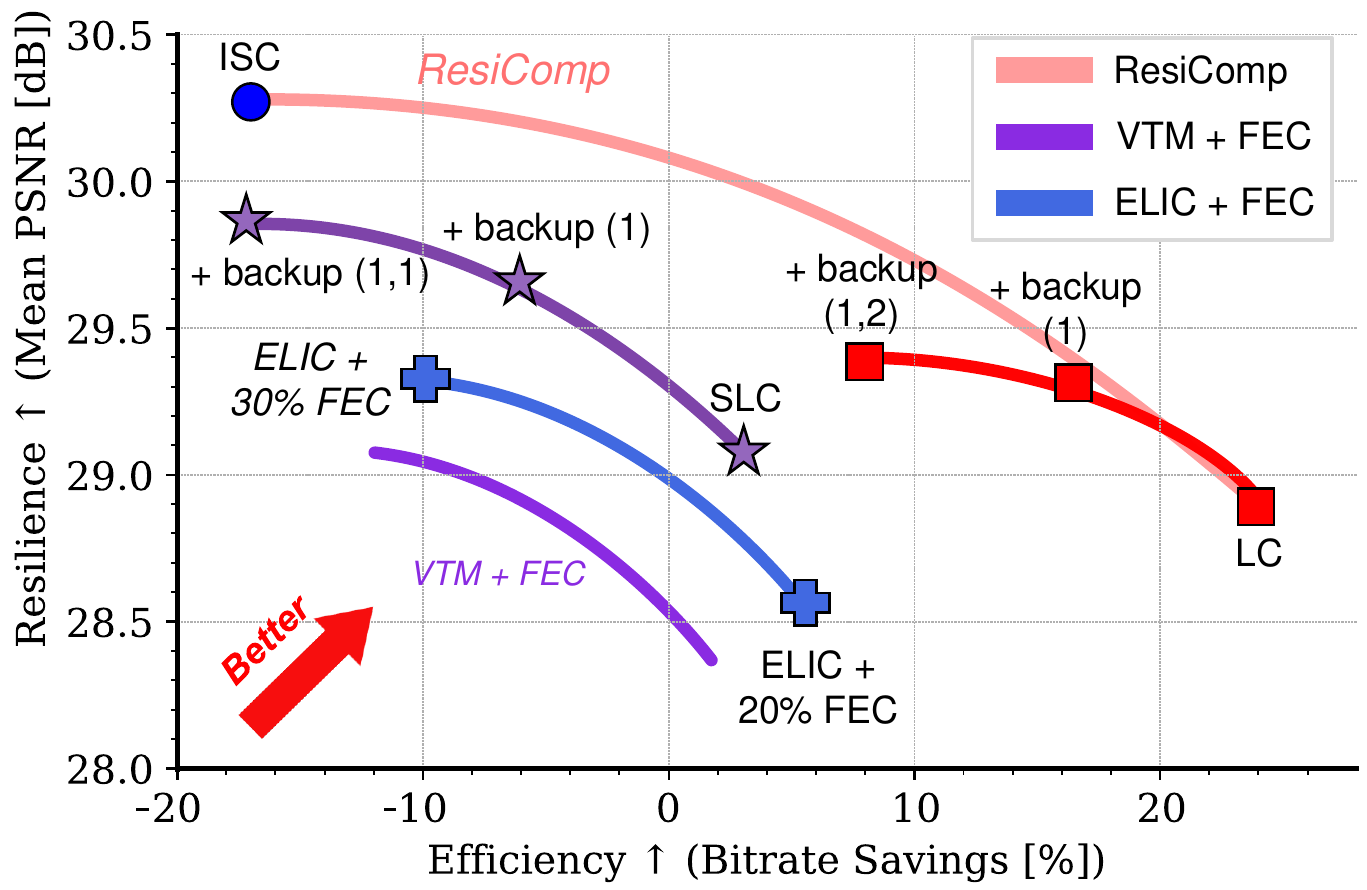}
    	\caption{Efficiency-resilience trade-off on Kodak. We provide backup packets to protect the important layers for our SLC and LC schemes, where list in parentheses is the index of the package being backed up. }
    	\label{Fig_ablation_benchmark}
    	\vspace{0em}
    \end{figure}
    	
	\begin{figure}[t]
		\setlength{\abovecaptionskip}{0.cm}
		\setlength{\belowcaptionskip}{-0.cm}
		\centering
		{\includegraphics[width=0.47\textwidth]{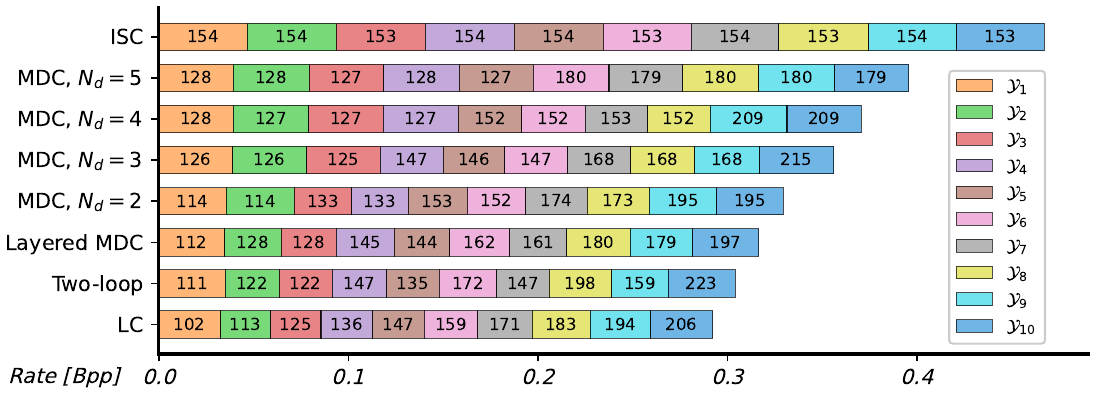}}
		\caption{The packet and slice size distribution across different context modes. In each row, distinct colors indicate separate token slices. The numerical value within each colored segment specifies the number of tokens in that slice, i.e., slice size. Additionally, the length of each colored sub-bar visually corresponds to the packet size, measured in bpp.}
		\label{Fig_packet_distribution}
	\end{figure}
	    
    \begin{figure*}[t]
    	\setlength{\abovecaptionskip}{0.cm}
    	\setlength{\belowcaptionskip}{-0.cm}
    	\begin{center}
    		\begin{minipage}{\textwidth}
    			\centering
    			{\includegraphics[width=0.85\textwidth]{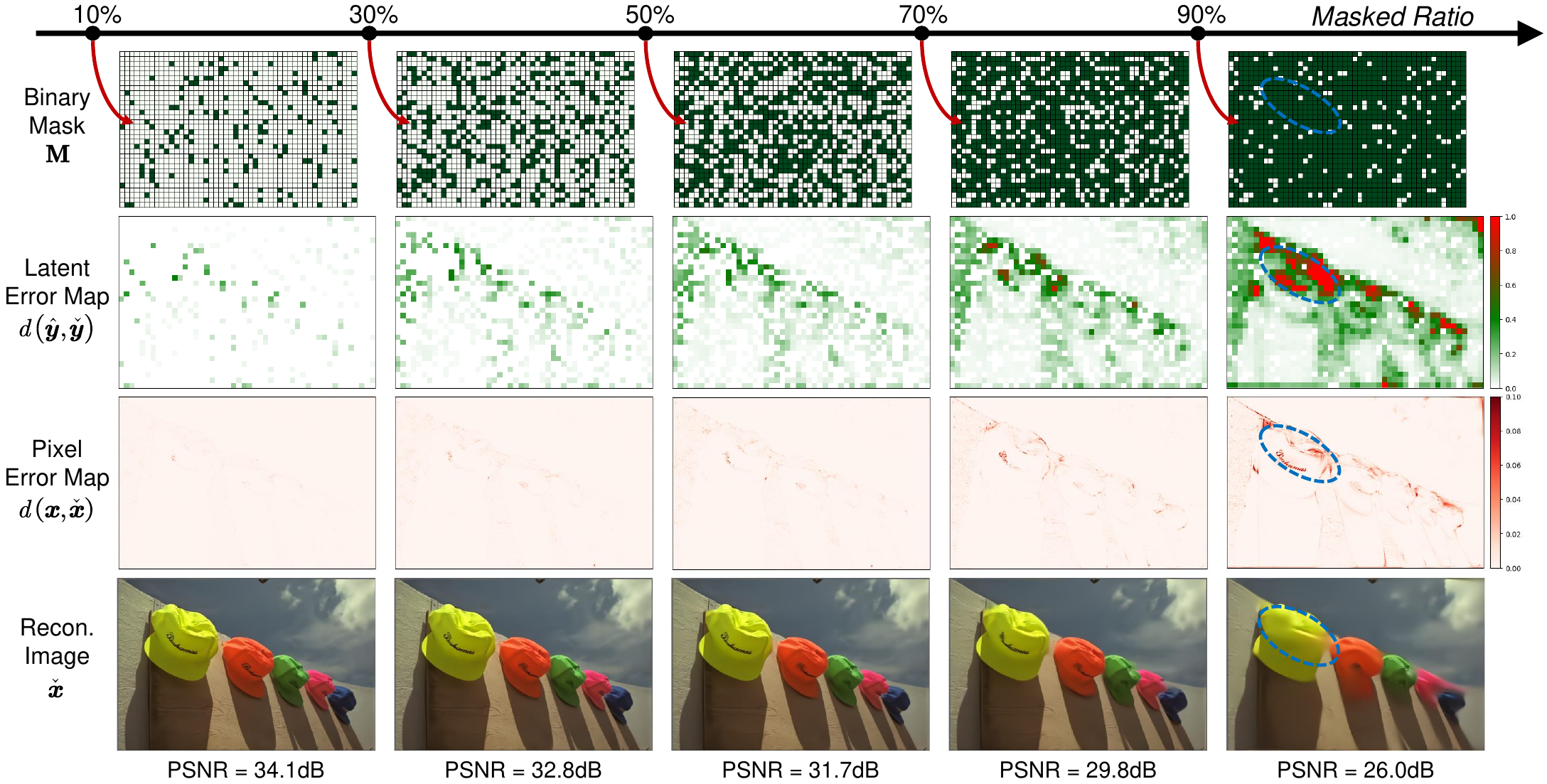}}
    		\end{minipage}
    		\caption{Qualitative evaluation of \emph{ResiComp}'s resilience to token loss: The latent tokens are masked with various ratios, gradually increasing from $10\%$ to $90\%$. The first row displays the binary mask $\mathbf{M}$, where the white and green grids represent the received and lost tokens, respectively. The second row shows the MSE error map between the concealed tokens $\checky$ and the encoded tokens $\haty$ in the latent space. The third and final rows display the reconstructed image $\checkx$ and its corresponding MSE error map. Please zoom in for a better view. }
    		\label{Fig_vis}
    		\vspace{-1em}
    	\end{center}
    \end{figure*}

	\begin{figure}[t]
		\setlength{\abovecaptionskip}{0.cm}
		\setlength{\belowcaptionskip}{-0.cm}
		\centering
		{\includegraphics[width=0.48\textwidth]{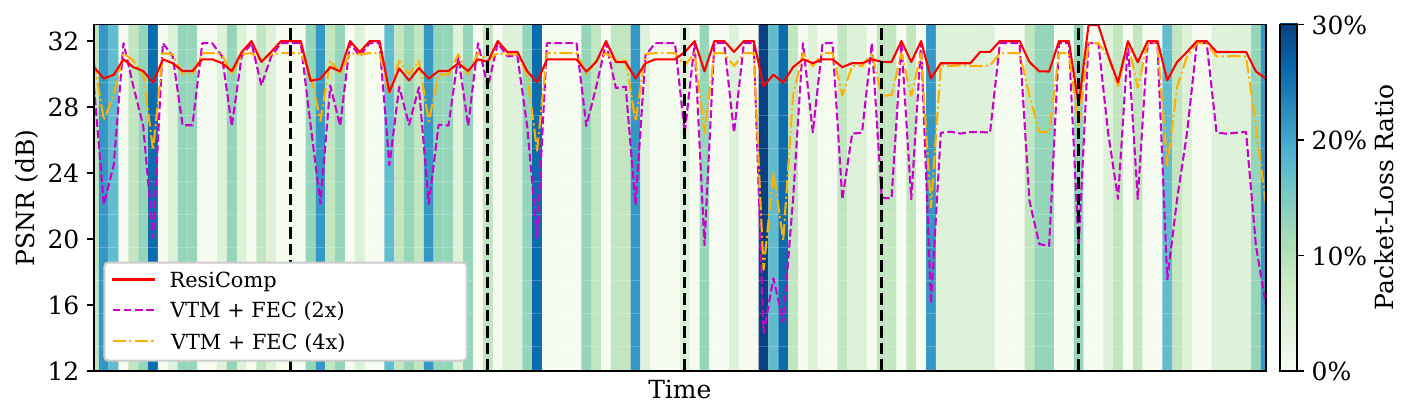}}
		\caption{Transmission under variable packet loss ratio: The gradation of background color indicates the instant packet loss ratio of current time slot. The black dashed line represents the granularity used to compute the average packet loss rate from the previous interval, which serves as evidence for updating the context mode in \emph{ResiComp} (or the FEC ratio for VTM) for the next interval. The legend "VTM + FEC ($n \times$)" indicates that the FEC ratio is selected as $n$ times the average packet loss rate.}
		\label{Fig_variable_network}
		\vspace{0em}
	\end{figure}

    As a widely used resilient technique, we also consider comparing with LC combined with unequal error protection (UEP).
    Specifically, we investigate two LC configurations: one with a protected base layer and an enhancement layer (SLC, as shown in Fig. \ref{Fig_scalable_layered_coding}), and another with several enhancement layers (illustrated in Fig. \ref{Fig_coding_mode}(b)).
    The efficiency-resilience performance for LC + UEP and SLC + UEP is depicted in Fig. \ref{Fig_ablation_benchmark}.
    It is evident that while UEP-based schemes can achieve varied efficiency-resilience trade-offs through increased redundancy, their overall performance falls short of ResiComp.
    This finding verifies our choice to against packet loss via changing the coding structure rather than increasing packet redundancy.

	{
	\subsubsection{Comparing with NIC + FEC}
	To facilitate a more straightforward comparison, we also include the ELIC + FEC scheme in Fig. \ref{Fig_ablation_benchmark}. It is evident that the proposed \emph{ResiComp} outperforms ELIC + FEC by a large margin. Since the $g_a$ and $g_s$ architectures of ELIC and \emph{ResiComp} are identical, this result demonstrates the dual-functional MVTM module provides a clear performance gain in packet-lossy channels.}

	\subsubsection{Rate Distribution}

	In Fig. \ref{Fig_packet_distribution}, we demonstrate the distribution of packet and slice sizes under different context modes.
	The factor $\beta$ is adjusted to ensure that packet sizes are comparable.
	Observing the rate distribution across each context mode reveals that latent slices with more contexts encode a larger number of latents within a similar bit budget, which verifies the flexibility of proposed approach.

	\subsubsection{Performance under Lossy Network with Variable Loss Ratios}
	
	In Fig. \ref{Fig_variable_network}, we evaluate a packet-lossy network with highly variable packet loss ratios by randomly alternating between Markov packet loss models EP3 and EP5. Both \emph{ResiComp} and the VTM + FEC scheme dynamically adjust their context mode and FEC ratio based on the packet loss history from the previous interval. The results clearly demonstrate that \emph{ResiComp} maintains stable reconstruction quality despite significant fluctuations in packet loss rates. In contrast, VTM + FEC encounters challenges: excessive redundancy reduces compression efficiency, while inadequate protection leads to frequent decoding failures.

	\subsection{Computational Complexity Comparison}

	\begin{table}[t]
		\centering
		\caption{
			Computational complexity, averaged encoding/decoding latency, and BD-Rate (\%) w.r.t. BPG comparison
		}
		\tabcolsep=0.2cm
		\resizebox{\linewidth}{!}{
			\begin{threeparttable}
				
				\begin{tabular}{m{1.7cm}<{}m{0.4cm}<{}m{1.2cm}<{\centering}m{1.2cm}<{\centering}m{1.2cm}<{\centering}m{1cm}<{\centering}m{0.8cm}<{\centering}}
					\toprule
					\multirow{2}{*}{Method} & \multirow{2}{*}{$L$} & \multicolumn{3}{c}{Inference Time $^*$} & \multirow{2}{*}{BD-rate} & \multirow{2}{*}{Param.} \\
					\cmidrule(lr){3-5}
					&& {GPU Enc.} & {GPU Dec.} & {CPU Dec.} & (\%) \\
					\midrule
					JointAR \cite{minnen2018} & $N$ & $46$ms &  $>10^3$ms & $1.5$s & $-10.64$ & $14$M \\
					GMM-Attn \cite{cheng2020learned} & $N$ & $78$ms & $>10^3$ms & $5.9$s & $-17.43$ & $20$M \\
					ELIC \cite{he2022elic, ELICReimplementation} & $10$ & $255$ms & $135$ms & 2.6s & $-26.85$ & $34$M \\
					MLIC \cite{jiang2023mlic} & $6$ & $220$ms & $170$ms & / & $-29.82$ & $116$M \\
					TCM \cite{liu2023learned} & $10$ & $217$ms & $162$ms & 2.2s & $-30.41$ & $45$M \\
					\emph{ResiComp} & $[1, N]$\tnote{\dag} & $251$ms & $232$ms & 3.6s & $-27.90$ & $128$M \\
					\bottomrule
				\end{tabular}
				
				\begin{tablenotes}
					\scriptsize
					\item[*] All experiments are conducted on the same platform with an Intel Xeon Gold 6226R CPU, an RTX 3090 GPU, with PyTorch 2.1.0 and CUDA 11.8. For ELIC, we use the re-implemented version from \cite{ELICReimplementation}. For MLIC, we take the speed and RD data from their paper  \cite{jiang2023mlic}.
					\item[\tnote{\dag}] $[1, N]$ means \emph{ResiComp} supports any slice number $L$ between $1$ to $N$. Here, we use the LC mode with $L=10$ slices for inference time and BD-rate evaluation.
				\end{tablenotes}
				\vspace{-.5em}
			\end{threeparttable}
		}
		\label{tab:complexity}
	\end{table}
	
	We report the computational complexities and BD-rates \cite{bjontegaard2001calculation} of \emph{ResiComp} compared to existing NICs in Table \ref{tab:complexity}. 
	In the latency test, we measured the encoding and decoding times on a GPU, as well as the decoding time on a CPU for a Kodak image. 
	The BD-rate \cite{bjontegaard2001calculation} represents the compression efficiency (with BPG as the anchor, smaller is better), calculated under ideal conditions without packet loss.
	We also report $L$ (the number of slices into which $\haty$ is divided for context modeling) in Table \ref{tab:complexity}. 
	For example, ELIC has $10$ slices, consisting of $5$ channel-slices, each multiplied by $2$ spatial-slices (anchor part and non-anchor part).
	Unlike existing NICs which use a fixed slice number and context mode,  \emph{ResiComp} features a flexible context model that supports completely customizable slice partition schedules and context construction.	These results demonstrate that \emph{ResiComp} achieves top-tier compression efficiency, with a reasonable number of parameters, although it runs slightly slower than other competitors with more complex entropy models.
	
	Specifically, running the Transformer at $512$x$768$ image once on CPU takes about $t_1=0.35$s. Running the decoder $g_s$ on CPU requires about $t_2=0.63$s. 
	The total CPU decoding time can be approximate calculated as $K_t \times t_1 + t_2$, where $K_t$ denotes the number of Transformer inference iterations.  Since inference from full mask tokens can be cached, we have $K_t=L-1$ for LC mode, $\lfloor L / N_d \rfloor - 1$ for MDC mode, and $0$ for ISC mode. Additionally, in the presence of packet loss, running the latent PLC module requires executing the Transformer once more.
	We emphasize that our MVTM-based Transformer is elegantly simple yet powerful, which will benefit from future	research into speeding up transformers.

    \subsection{Visualization}

	In Fig. \ref{Fig_vis}, we present an intuitive visualization of \emph{ResiComp}'s resilience to token loss with different mask ratios. From the latent and pixel error maps, we observe that \emph{ResiComp} is capable of concealing the majority of lost tokens, with predicted values close to $\haty$ (marked in light green). 
	In the extreme case, with only $10\%$ of tokens received, we observe that \emph{ResiComp} produces some unfaithful reconstructed tokens (marked in red in the latent error map). 
	Despite some loss of detailed textures, we emphasize that \emph{ResiComp} successfully prevents a collapse in overall reconstruction quality and still retains promising reconstructions with faithful structure.

    \section{Conclusion, Limitations, and Future Works}\label{section_conclusion}

	We have proposed \emph{ResiComp}, a novel loss-resilient neural image transmission framework that elevate the resilience of NICs while maintains top-tier compression efficiency.
	The key of \emph{ResiComp} is to unify the entropy modeling and latent PLC task in a simple and effective approach focused on context modeling.
	We have showed that our dual-functional Transformers, trained on masked visual token modeling, exhibit powerful and flexible context modeling ability.
	Furthermore, we have proposed diverse context modes to explicitly organize the contextual dependencies across token slices, which enables our \emph{ResiComp} to achieve multiple efficiency-resilience trade-offs within a single model.
	Extensive experiments have demonstrated \emph{ResiComp} outperforms redundancy-based resilient codecs, such as VTM + FEC, offering superior efficiency-resilience trade-offs. 

	While \emph{ResiComp} demonstrate how to elevate the resilience of NICs, its generalization capability is not always satisfactory, particularly for domain-specific images such as screen content, comics, and medical images, which is a common challenge for NICs trained on natural image datasets.
	Our future work will focus on extending \emph{ResiComp} to inter-frame coding in video communication and the in-depth integration with physical layer technologies to further improve the compression efficiency and loss resilience.
	We hope this research can inspire the development of neural codecs for RTC applications, such as online meetings, live video streaming, cloud gaming, and other applications.

    \ifCLASSOPTIONcaptionsoff
    \newpage
    \fi

    \bibliographystyle{IEEEtran}

    \bibliography{Ref}

\begin{IEEEbiography}[{\includegraphics[width=1in,height=1.25in,clip,keepaspectratio]{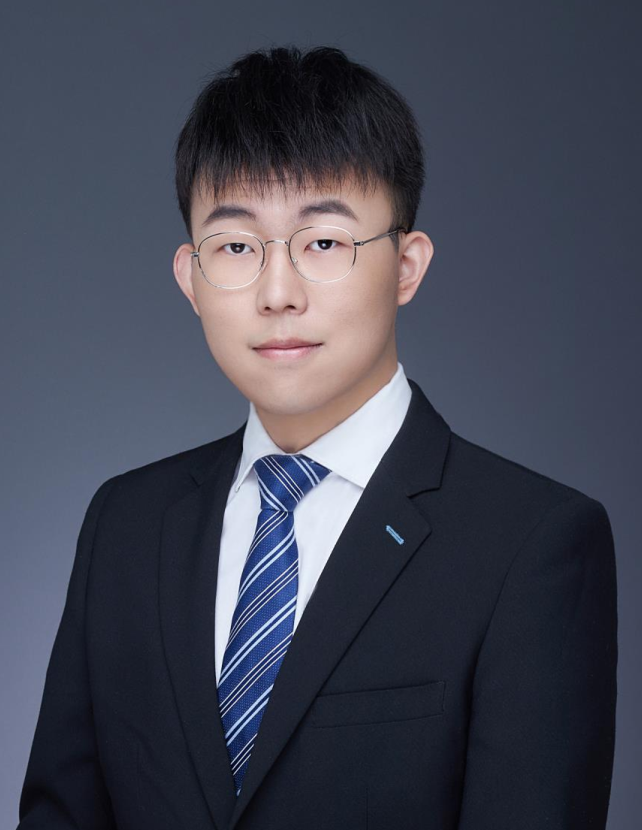}}]{Sixian Wang}
	received his B.S. degree from the Beijing University of Posts and Telecommunications, Beijing, China. He is currently pursuing the Ph.D. degree with Beijing University of Posts and Telecommunications. His research focuses on source and channel coding, computer vision, and intelligent communications. He has been selected for the China Association for Science and Technology’s (CAST) “Young Elite Scientists Sponsorship Ph.D. Program”.
\end{IEEEbiography}

\begin{IEEEbiography}[{\includegraphics[width=1in,height=1.25in,clip,keepaspectratio]{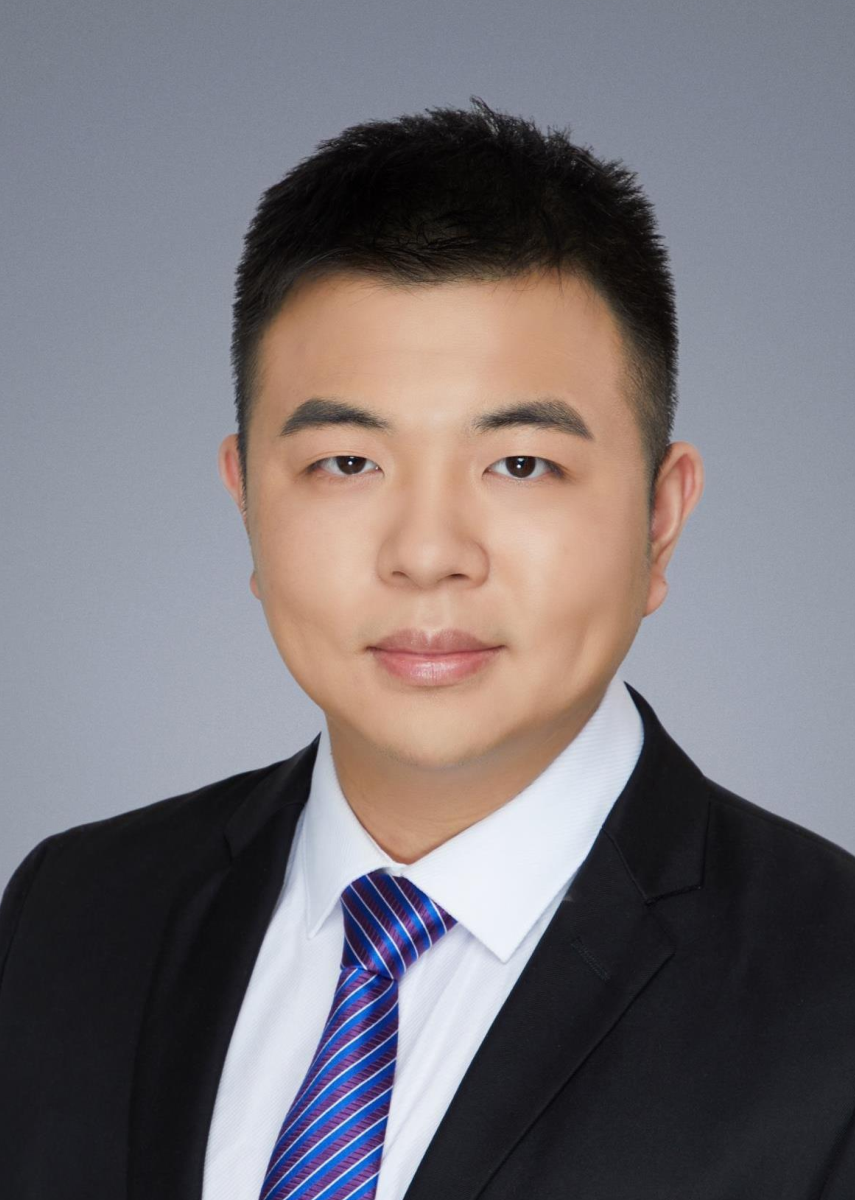}}]{Jincheng Dai}
	received his B.S. and Ph.D. degrees from Beijing University of Posts and Telecommunications, Beijing, China. He is currently an Associate Professor with the Key Laboratory of Universal Wireless Communications, Ministry of Education, Beijing University of Posts and Telecommunications. His research focuses on source and channel coding, generative AI, physical AI, and intelligent communications. He has been selected for the China Association for Science and Technology’s (CAST) “Young Elite Scientists Sponsorship Program” and the Xiaomi Young Scholars Program. As a core contributor, he has received the China Electronics Society’s First-Class Natural Science Award. He has received several accolades, such as the Excellent Science and Technology Paper Award from CAST.
\end{IEEEbiography}

\begin{IEEEbiography}[{\includegraphics[width=1in,height=1.25in,clip,keepaspectratio]{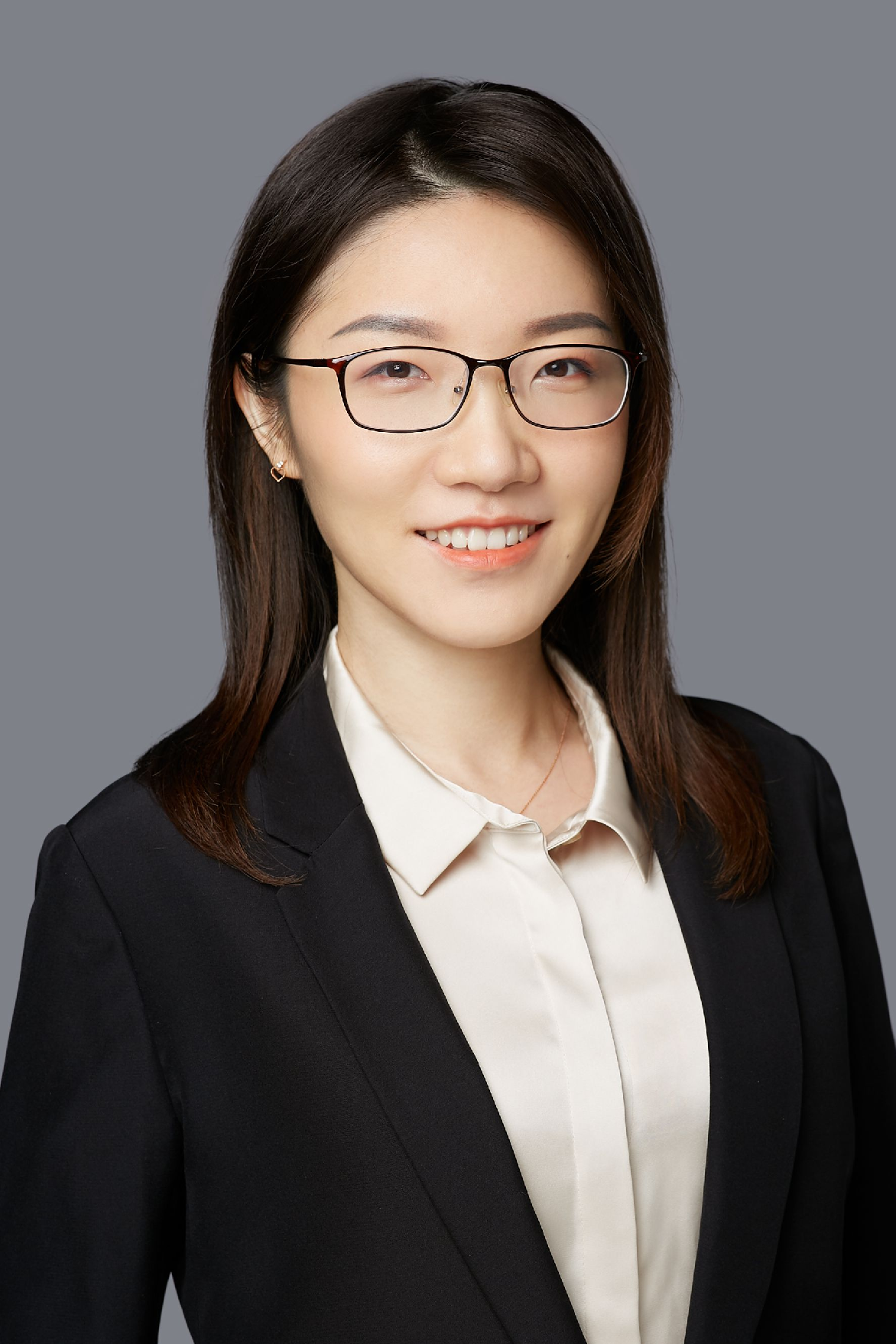}}]{XiaoQi Qin}	
	received her B.S., M.S., and Ph.D. degrees from Electrical and Computer Engineering with Virginia Tech. She is currently an Associate Professor of School of Information and Communication Engineering with Beijing University of Posts and Telecommunication. Her research focuses on exploring performance limits of next-generation wireless networks, and developing innovative solutions for intelligent and efficient machine-type communications.
\end{IEEEbiography}

\begin{IEEEbiography}[{\includegraphics[width=1in,height=1.25in,clip,keepaspectratio]{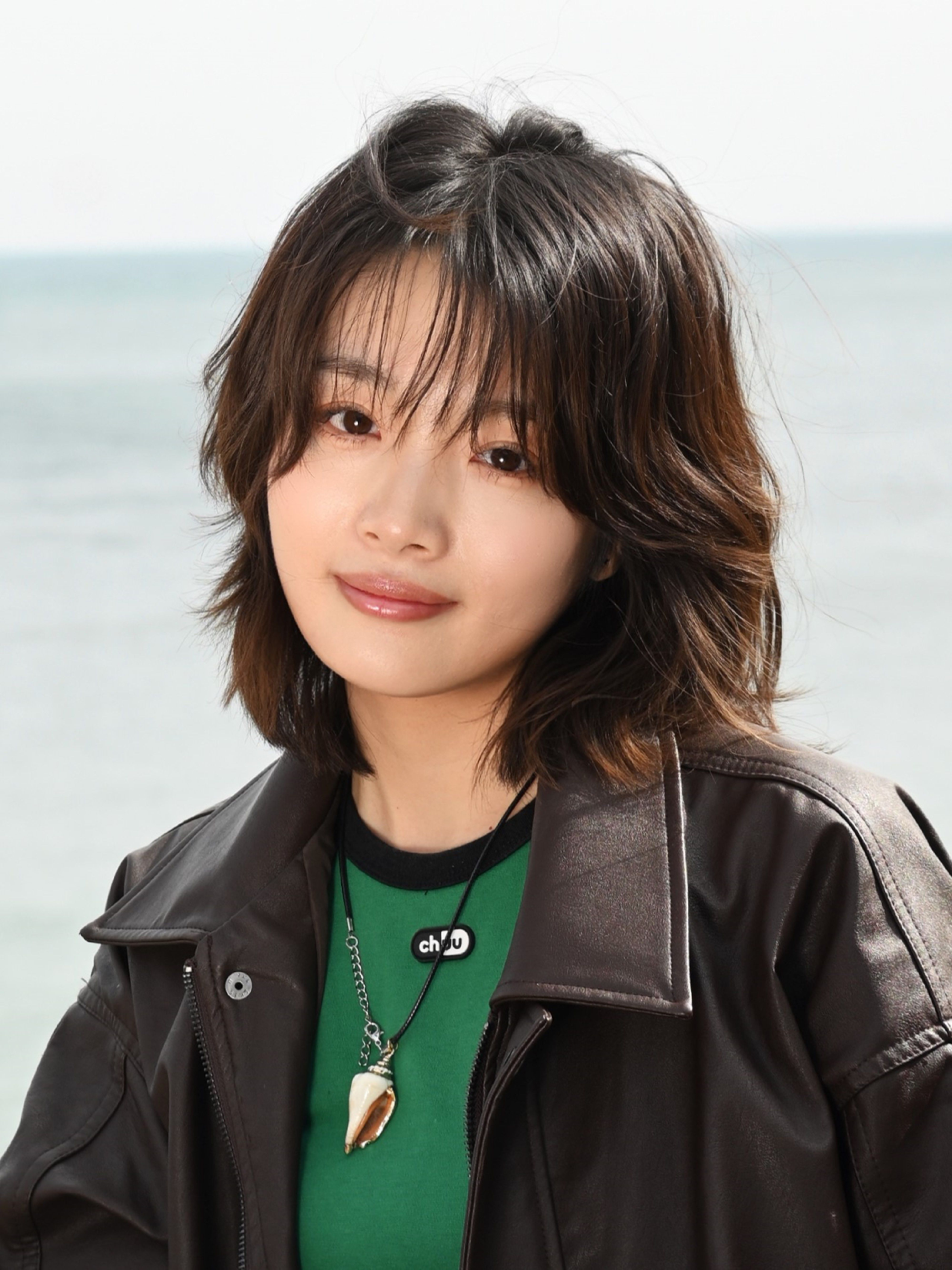}}]{Ke Yang}
	received her B.S. degrees from Beijing University of Posts and Telecommunications, Beijing, China. She is currently pursuing the Ph.D. degree with Beijing University of Posts and Telecommunications. Her research focuses on source and channel coding and intelligent communications.
\end{IEEEbiography}

\begin{IEEEbiography}[{\includegraphics[width=1in,height=1.25in,clip,keepaspectratio]{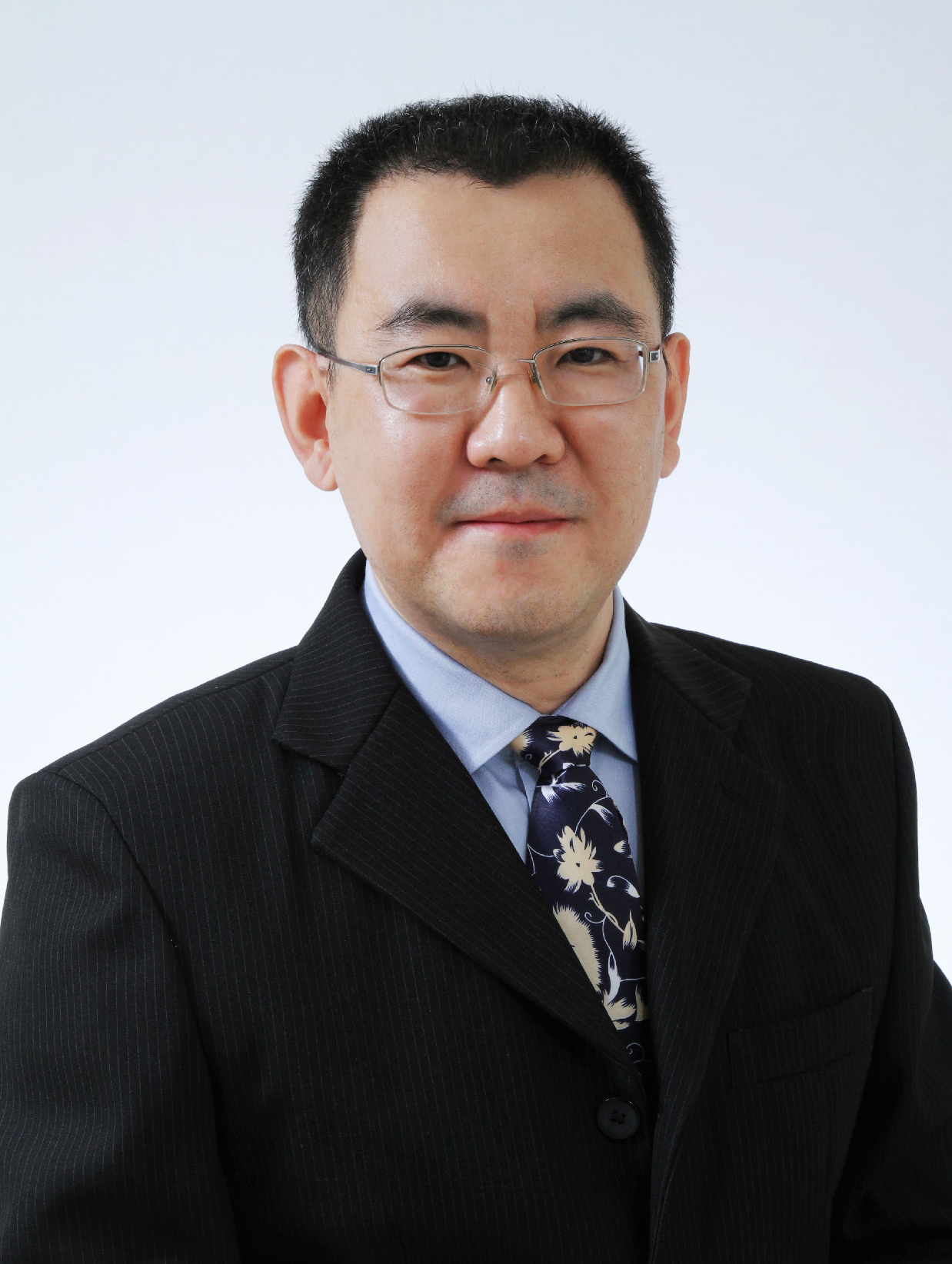}}]{Kai Niu}
	received his B.S. and Ph.D. degrees from Beijing University of Posts and Telecommunications, Beijing, China. He is currently a Professor with Beijing University of Posts and Telecommunications. Since 2008, he has been a Senior Member with the Chinese Institute of Electronics (CIE) and a committee member of the Information Theory Chapter of CIE. He currently serves as an Associate Editor with the IEEE Communications Letters. His research interests include coding theory and its applications.
\end{IEEEbiography}

\begin{IEEEbiography}[{\includegraphics[width=1in,height=1.25in,clip,keepaspectratio]{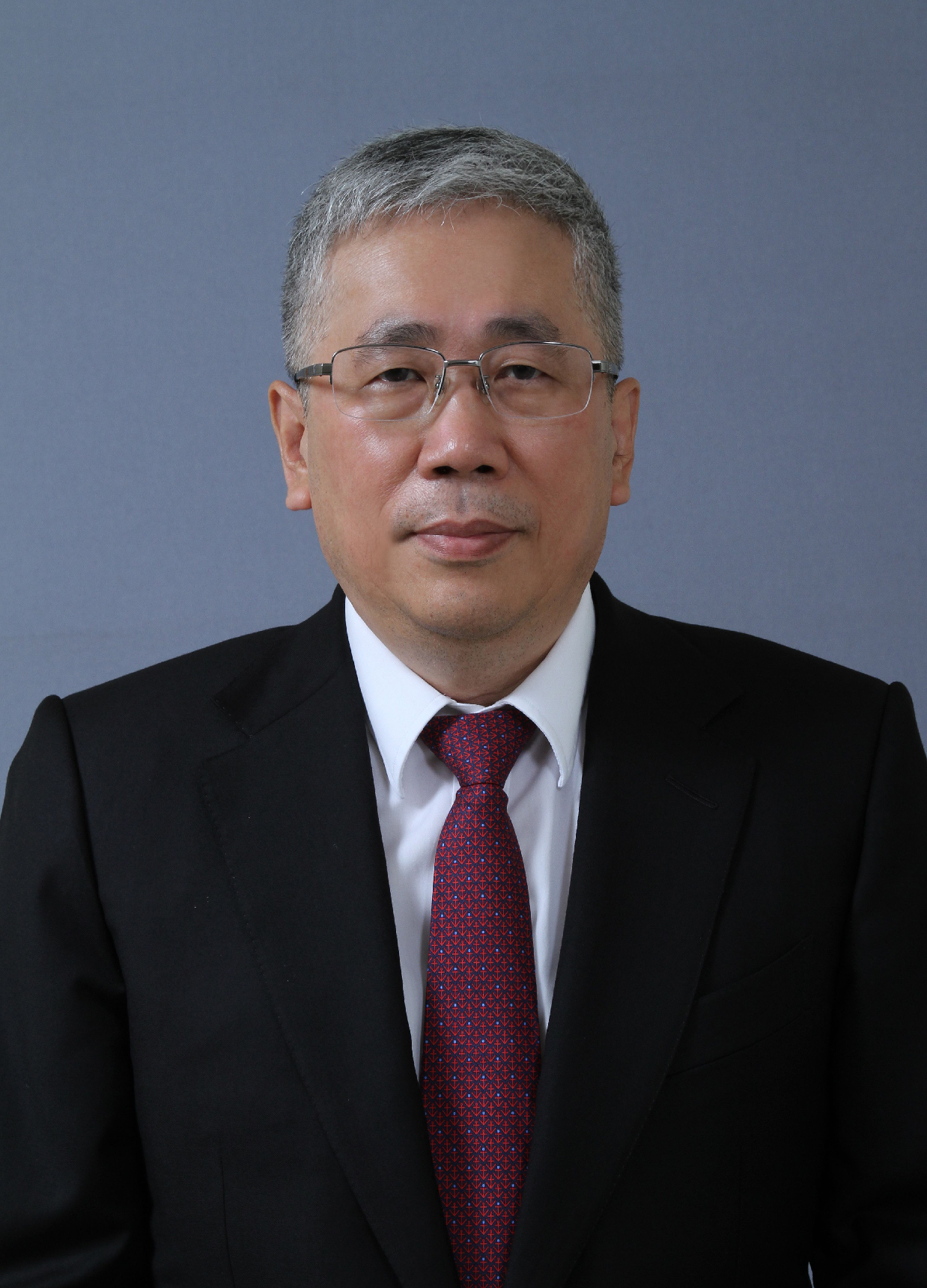}}]{Ping Zhang}
	is currently a professor of School of Information and Communication Engineering at Beijing University of Posts and Telecommunications, the director of State Key Laboratory of Networking and Switching Technology, a member of IMT-2020 (5G) Experts Panel, a member of Experts Panel for China's 6G development. He served as Chief Scientist of National Basic Research Program (973 Program), an expert in Information Technology Division of National High-tech R\&D program (863 Program), and a member of Consultant Committee on International Cooperation of National Natural Science Foundation of China. His research interests mainly focus on wireless communication. He is an Academician of the Chinese Academy of Engineering (CAE).
\end{IEEEbiography}

\end{document}